\documentclass[aps,prd,twocolumn,superscriptaddress,nofootinbib,amssymb]{revtex4-2}
\usepackage{amsmath,amsfonts}
\usepackage{graphicx}
\usepackage[caption=false]{subfig}
\usepackage[dvipsnames,usenames]{xcolor}
\usepackage{listings}
\usepackage[colorlinks,citecolor=blue]{hyperref}
\usepackage{ulem}
\usepackage{orcidlink}

\begin{document}

\title{Evaluating approximate asymptotic distributions for fast neutrino flavor conversions in a periodic 1D box}

\author{Zewei Xiong\orcidlink{0000-0002-2385-6771}}
\email[Email: ]{z.xiong@gsi.de}
\affiliation{GSI Helmholtzzentrum {f\"ur} Schwerionenforschung, Planckstra{\ss}e 1, 64291 Darmstadt, Germany}

\author{Meng-Ru Wu\orcidlink{0000-0003-4960-8706}}
\affiliation{Institute of Physics, Academia Sinica, Taipei 11529, Taiwan}
\affiliation{Institute of Astronomy and Astrophysics, Academia Sinica, Taipei 10617, Taiwan}
\affiliation{Physics Division, National Center for Theoretical Sciences, Taipei 10617, Taiwan}

\author{Sajad Abbar\orcidlink{0000-0001-8276-997X}}
\affiliation{Max-Planck-Institut f\"ur Physik (Werner-Heisenberg-Institut), F\"ohringer Ring 6, D-80805 M\"unchen, Germany}

\author{Soumya Bhattacharyya\orcidlink{0000-0001-5331-4597}}
\affiliation{Institute of Physics, Academia Sinica, Taipei 11529, Taiwan}

\author{Manu George\orcidlink{0000-0002-8974-5986}}
\affiliation{Institute of Physics, Academia Sinica, Taipei 11529, Taiwan}

\author{Chun-Yu Lin\orcidlink{0000-0002-7489-7418}}
\affiliation{National Center for High-performance Computing, National Applied Research Laboratories, Hsinchu 30076, Taiwan}

\date{\today}

\begin{abstract}
The fast flavor conversions (FFCs) of neutrinos generally exist in core-collapse supernovae and binary neutron-star merger remnants and can significantly
change the flavor composition and affect the dynamics and nucleosynthesis processes.
Several analytical prescriptions were proposed recently to approximately explain or predict the asymptotic outcome of FFCs for systems with different initial or boundary conditions, with the aim for providing better understandings of FFCs and for practical implementation of FFCs in hydrodynamic modeling.
In this work, we obtain the asymptotic survival probability distributions of FFCs in a survey over thousands of randomly sampled initial angular distributions by means of numerical simulations in one-dimensional boxes with the periodic boundary condition. 
We also propose improved prescriptions that guarantee the continuity of the angular distributions after FFCs.
Detailed comparisons and evaluation of all these prescriptions with our numerical survey results are performed.
The survey dataset is made publicly available to inspire the exploration and design for more effective methods applicable to realistic hydrodynamic simulations.
\end{abstract}

\maketitle
\graphicspath{{./figures/}}

\section{Introduction}
\label{sec:introduction}
A great amount of neutrinos are produced in dense astrophysical environments such as core-collapse supernovae (CCSNe) and the remnants of binary neutron-star mergers (BNSMs).
Their fluxes are so intense that the coherent forward scattering among those neutrinos can lead to significant changes in their flavor content through the collective flavor instabilities, particularly the fast flavor conversion (FFC; see e.g., \cite{duan2010collective,mirizzi2016supernova,tamborra2021new,richers2022fast,capozzi2022neutrino,volpe2023neutrinos} for reviews) at the vicinity of the cores of CCSNe and accretion disks of BNSMs, which can play important roles in the dynamics and the nucleosynthesis of those environments \cite{wu2017imprints,stapleford2020coupling,xiong2020potential,george2020fast,li2021neutrino,just2022fast,fernandez2023fast,fujimoto2023explosive,nagakura2023roles,ehring2023fast,ehring2023fast2}.

FFC happens when the angular distribution of the neutrino lepton number between any two distinct flavors takes both positive and negative values simultaneously \cite{morinaga2022fast} with the transition points often dubbed as ``zero crossings.''
Given that the multidimensional simulations usually provide only the angular moments instead of the full distributional information, various approximate or parametric methods were adopted \cite{dasgupta2018simple,abbar2019fast,abbar2020searching,glas2020fast,johns2021fast,nagakura2021new,richers2022evaluating,abbar2023applications} and found the existence of FFC near or even inside the neutrinosphere in CCSNe \cite{tamborra2017flavor,abbar2019occurrence,azari2019linear,azari2020fast,abbar2020fast,nagakura2019fast,glas2020fast,nagakura2021occurrence,abbar2021characteristics,morinaga2020fast,harada2022prospects} as well as ubiquitously in the postmerger remnants of BNSMs \cite{wu2017fast,richers2022evaluating}.

The spatial and temporal scales associated with the development of the fast flavor instability can be in subcentimeter and subnanosecond, much shorter than the typical scales considered in the hydrodynamical simulations for CCSNe and BNSMs.
This naturally brings up a challenge to incorporate FFC into the hydrodynamical simulations.
To overcome this challenge, one possible solution is to decompose this problem into two scale hierarchies: performing the local dynamical simulations at a small scale and summarizing with useful parametric prescriptions that can be applied to the hydrodynamic simulation more efficiently.

The outcome of FFCs has been extensively studied based on local dynamical simulations in tiny boxes with a periodic boundary condition \cite{martin2020dynamic,bhattacharyya2020late,bhattacharyya2021fast,wu2021collective,richers2021neutrino,zaizen2021nonlinear,richers2021particle,bhattacharyya2022elaborating,grohs2022neutrino,abbar2022suppression,richers2022code,zaizen2023simple} and may be affected by adopting a different boundary condition \cite{zaizen2023characterizing}.
These studies suggest that the flavor conversions undergo the kinematic decoherence in general \cite{raffelt2007self,abbar2019fast,johns2020fast,bhattacharyya2021fast,xiong2023symmetry} and reach asymptotically to quasistationary states achieving complete or partially flavor equilibration as allowed by the conservation of neutrino lepton number \cite{raffelt2007self,duan2009symmetries}, when coarse grained over the box size.

An early attempt to obtain an analytical description on the asymptotic distribution was made in a homogeneous neutrino gas \cite{xiong2021stationary}, and a growing number of schemes have been recently proposed for the local simulations allowing the advection of neutrinos inside the box \cite{bhattacharyya2021fast,bhattacharyya2022elaborating,zaizen2023simple}.
Those methods often contain an artificial discontinuity for the survival probability distribution near the zero crossing, which is further passed to the asymptotic angular distributions of neutrino number densities.
In this paper, we propose new prescriptions by imposing a continuous transition at the zero crossing and performed numerical FFC simulations for $\sim \mathcal O(8000)$ systems with randomly sampled initial angular distributions that cover the major parameter space for FFCs to occur near the neutrino decoupling regions. 
For each prescription, we compare the predicted asymptotic angular distributions after FFC with those obtained by numerical simulations and evaluate in detail their performance using different metrics of errors.
Our improved prescriptions of the asymptotic distributions not only predict the angular moments in the asymptotic state more accurately, but also can be directly implemented in the discrete-ordinate neutrino transport with the advection on a large scale
\cite{shalgar2022supernova,nagakura2022time,shalgar2023neutrino,xiong2023evolution,nagakura2023connecting}.

We describe the simulation setup over thousands of parameter sets in Sec.~\ref{sec:simulations}.
We present various analytical prescriptions to determine the asymptotic distributions with and without continuous transitions at the zero crossing in Sec.~\ref{sec:prescriptions}.
The results of the performance evaluation for those asymptotic prescriptions are presented in Sec.~\ref{sec:results}.
Finally, we provide our discussions and conclusions in Sec.~\ref{sec:conclusions}.
We adopt natural units ($\hbar=c=1$) throughout the paper.

\section{Survey of simulations}
\label{sec:simulations}
\subsection{Equation of motion}
We use the code \textsc{cose$\nu$} \cite{george2023cosenu} to evolve the FFCs in a similar setup of one-dimensional (1D) box as described in Ref.~\cite{wu2021collective} assuming translation symmetry in the $x$ and $y$ directions, axial symmetry around the $z$ axis, and periodic boundary condition in the $z$ direction.
We consider in the simulation that the oscillations start from electron flavor $\nu_e$ ($\bar\nu_e$) initially and can be converted to one heavy-lepton flavor $\nu_x$ ($\bar \nu_x$).\footnote{The effects of heavy-lepton flavor neutrinos in the initial condition will be discussed in Sec.~\ref{sec:nux}.}
We neglect the vacuum mixing and neutrino-matter forward scattering.
Although it is reported that the so-called collisional flavor instability induced by neutrino emission and absorption may interplay with the FFC \cite{johns2023collisional,johns2022collisional,padilla2022neutrino2,lin2023collision,xiong2022collisional,kato2023flavor,liu2023systematic}, we neglect all collisional processes in the 1D-box setup that we consider.
The equation of motion (EOM) for the normalized neutrino (antineutrino) density matrix $\varrho$ ($\bar\varrho$) is given by
\begin{align}
    (\partial_t+v_z \partial_z) \varrho(t,z,v_z) & = -i[H(t,z,v_z), \varrho(t,z,v_z)], \nonumber\\
    (\partial_t+v_z \partial_z) \bar\varrho(t,z,v_z) & = i[H^*(t,z,v_z), \bar\varrho(t,z,v_z)],
    \label{eq:eom}
\end{align}
with the Hamiltonian of coherent forward scattering at specific $t$ and $z$
\begin{align}
    H(v_z) = & \mu \int_{-1}^1 dv_z' (1-v_z v_z') [g_\nu(v_z')\varrho(v_z') -g_{\bar\nu}(v_z')\bar \varrho^*(v_z')]
    \label{eq:eom_H}
\end{align}
where $\mu=\sqrt{2} G_F n_{\nu_e}$, $G_F$ is the Fermi constant, $n_{\nu_e}$ is the number density for $\nu_e$, and  $g_{\nu}$ ($g_{\bar\nu}$) is the initial angular distribution for $\nu_e$ ($\bar\nu_e$) as a function of the projected velocity $v_z$.
The distribution $g_\nu$ for $\nu_e$ is normalized with the zeroth moment $I_\nu=\int_{-1}^1 dv_z g_\nu(v_z)=1$.
The zeroth moment for $\bar\nu_e$, $I_{\bar\nu}=\int_{-1}^1 dv_z g_{\bar\nu}(v_z)=n_{\bar \nu_e}/n_{\nu_e}$, indicates whether the condition is $\nu_e$- or $\bar\nu_e$-dominant.
The angular distribution of neutrino electron lepton number ($\nu$ELN)
\begin{equation}
    G(v_z)=g_{\nu}(v_z)-g_{\bar\nu}(v_z),
\end{equation}
determines the existence of fast flavor instability.

Without explicitly including the vacuum mixing in the EOM, we trigger the FFC by seeding random perturbations in the initial condition
\begin{align}
    \varrho_{ee}(z,v_z) = \bar\varrho_{ee}(z,v_z) = & [1+\sqrt{1-\epsilon^2(z)}]/2, \nonumber\\
    \varrho_{xx}(z,v_z) = \bar\varrho_{xx}(z,v_z) = & [1-\sqrt{1-\epsilon^2(z)}]/2, \nonumber\\
    \varrho_{ex}(z,v_z) = \bar\varrho_{ex}(z,v_z) = & \epsilon(z)/2,
\end{align}
where a real number $\epsilon$ is randomly assigned for each $z$ and follows a uniform distribution between 0 and $10^{-2}$.

\subsection{Setup and parameters}
We consider two types of initial angular distributions.
The first one is described by a Gaussian function \cite{yi2019dispersion}
\begin{equation}
    g_{\nu(\bar \nu)}(v_z)\propto \exp[-(v_z-1)^2/(2\sigma^2_{\nu (\bar\nu)})],
\end{equation}
and the second one is obtained from maximum-entropy closure \cite{richers2022evaluating}
\begin{equation}
    g_{\nu(\bar \nu)}(v_z)\propto \exp[v_z/\sigma_{\nu (\bar\nu)}].
\end{equation}
For both types, $\sigma_{\nu (\bar\nu)}$ is a parameter associated with the neutrino (antineutrino) flux factor, i.e., the ratio between the first and zeroth angular moments,
\begin{equation}
    F_{\nu (\bar\nu)} = \frac{J_{\nu (\bar\nu)}}{I_{\nu (\bar\nu)}}
\end{equation}
where $J_{\nu (\bar\nu)} = \int_{-1}^1 dv_z v_z g_{\nu (\bar\nu)}(v_z)$.
For both types, since the zeroth moment of $\nu_e$ is normalized, the initial angular distributions can be uniquely determined by three parameters $I_{\bar\nu}$, $F_{\nu}$, and $F_{\bar\nu}$.

We choose these three parameters in the following way.
First we randomly assign $I_{\bar\nu}$ and $F_{\bar\nu}$ following uniform distributions ranging from 0.5 to 1.6 and from 0.3 to 0.9, respectively.
We take the lower limit for $F_{\bar\nu}$ as 0.3 for our survey. 
This is because a smaller flux factor implies more isotropic neutrino angular distribution generally obtained near the neutrino optically thick region associated with higher density and temperature.
Under those conditions, the highly degenerate electrons leads to a large neutrino chemical potential of electron flavor so that $g_{\nu}$ dominates over $g_{\bar\nu}$  in the whole $v_z$ range, i.e., the fast flavor instability is less likely to occur for $F_{\bar\nu}\lesssim 0.3$.
We also do not consider flux factor higher than 0.9 because neutrinos become highly collimated toward one direction and may not be well captured with the angular resolution in the current setup.
We then randomly assign $F_{\nu}$ in a uniform distribution from $\sim 0.65 F_{\bar\nu}$ to $\min(0.9, 1.6 F_{\bar\nu})$  where ``min'' stands for the minimum function. 
This constraint is to avoid the situation where either $\nu_e$ or $\bar\nu_e$ has much larger flux factor than the other, which usually does not occur in realistic systems because the decoupling regions of $\nu_e$ or $\bar\nu_e$ are not very far apart.

If a zero crossing at $v_c$ where $G(v_c)=0$ exists, we adopt this parameter set and perform the simulation in a periodic 1D box using the corresponding initial angular distributions.
Otherwise, this parameter set is rejected, and we continue to generate new parameters.
In all simulations, the number of spatial grids is $N_z=6000$.
The size of the 1D box is $L_z=1200 \mu^{-1}$.
We adopt the finite volume method as well as the seventh-order weighted essentially nonoscillatory scheme.

We repeat the same procedure above until 8000 parameter sets are adopted for the Gaussian-type distributions with zero crossings. 
In each parameter set we take two different angular resolutions with $N_{v_z}=50$ and $N_{v_z}=100$.
For the maximum-entropy distributions we use the same parameter sets of $I_{\bar\nu}$, $F_{\nu}$, and $F_{\bar\nu}$ as in the Gaussian type.
We further exclude those not having any zero crossings in the maximum-entropy type, which reduces the size of the sample to 7668 parameter sets for this case.

\subsection{Determination of the asymptotic distributions}
During the simulation of each set of initial angular distributions, the space-averaged survival probabilities for $\nu_e$ and $\bar\nu_e$ at each $t$ are defined as
\begin{align}
    \langle \varrho_{ee}\rangle_z(v_z) & = \frac{1}{L}\int dz \varrho_{ee}(z,v_z), \nonumber\\
    \langle \bar\varrho_{ee}\rangle_z(v_z) & = \frac{1}{L}\int dz \bar\varrho_{ee}(z,v_z),
\end{align}
respectively.
The overall space-averaged survival probabilities are
\begin{align}
    \langle P_{ee} \rangle =\int_{-1}^1 dv_z g_\nu(v_z) \langle \varrho_{ee}\rangle_z(v_z)\Big/\int_{-1}^1 dv_z g_\nu(v_z), \nonumber\\
    \langle P_{\bar e\bar e} \rangle =\int_{-1}^1 dv_z g_{\bar \nu}(v_z) \langle \bar\varrho_{ee}\rangle_z(v_z)\Big/\int_{-1}^1 dv_z g_{\bar\nu}(v_z),
\end{align}
respectively.

In the presence of the initial perturbation, both $\langle P_{ee} \rangle$ and $\langle P_{\bar e\bar e} \rangle$ start from values very close to 1 and decrease under the fast flavor instability until reaching the first minimum point.
They bounce back but do not return to 1.
Instead, they enter into a ringdown phase with gradually damped oscillation amplitude and eventually approach asymptotic values [see, e.g., Fig.~5(b) in Ref.~\cite{wu2021collective}].

We take a practical approach to determine whether the system has reached the asymptotic state as follows.
For each simulation, we record the times when $\langle P_{ee} \rangle$ reaches the first and second minima as $t_1$ and $t_2$, respectively and define $\Delta T=t_2-t_1$.
Then, we end the simulation at $t_f=t_2+N_t\Delta T$ with $N_t=20$ to cover roughly $N_t$ more periods during the ringdown phase.
Because $\langle \varrho_{ee}\rangle_z(v_z)$ and $\langle \bar\varrho_{ee}\rangle_z(v_z)$ may still fluctuate in time at the end of the simulation, we compute the time-averaged survival probabilities
\begin{align}
    P_{ee}(v_z) & = \frac{1}{\Delta T}\int_{t_f-\Delta T}^{t_f} dt \langle \varrho_{ee}\rangle_z(t,v_z), \nonumber\\
    P_{\bar e\bar e}(v_z) & = \frac{1}{\Delta T}\int_{t_f-\Delta T}^{t_f} dt \langle \bar\varrho_{ee}\rangle_z(t,v_z),
\end{align}
over the last time interval of $\Delta T$ as our final data outputs.

Since Eqs.~\eqref{eq:eom} and \eqref{eq:eom_H} imply the relation that $\varrho_{ee}(t,z,v_z)= \bar\varrho_{ee}(t,z,v_z)$ and $P_{ee}(v_z) = P_{\bar e\bar e}(v_z)$, 
the time- and space-averaged angular distributions after the FFCs can be computed as
$\tilde g_{\nu_e}(v_z) = g_\nu(v_z) P_{ee}(v_z)$ and  $\tilde g_{\bar\nu_e}(v_z) = g_{\bar\nu}(v_z) P_{ee}(v_z)$ accordingly. 
It follows that the zeroth and first moments for $\nu_e$ ($\bar\nu_e$) after the FFCs are $\tilde I_{\nu_e(\bar\nu_e)}=\int_{-1}^1 dv_z \tilde g_{\nu_e(\bar\nu_e)}$ and $\tilde J_{\nu_e(\bar\nu_e)}=\int_{-1}^1 dv_z v_z \tilde g_{\nu_e(\bar\nu_e)}$ respectively.
We store the time-averaged final distributions for the survival probability as well as the first two angular moments for $\nu_e$ and $\bar\nu_e$ described above for the entire sets with 8000 and 7668 different initial conditions for the Gaussian and maximum-entropy types, respectively.
The full dataset is available in \cite{dataset_fast_asymptotic}.

\section{Analytical prescriptions}
\label{sec:prescriptions}
For both Gaussian and maximum-entropy types that we consider, the initial $\nu$ELN distribution $G(v_z)$ is ensured to allow at most one zero crossing.
Thanks to this feature, we can divide the $v_z$ range into two parts separated by the zero crossing $v_c$.
The integrals over both parts are
\begin{align}
    I_+ & = \left| \int_{-1}^1 dv_z G(v_z) \Theta[G(v_z)] \right|, \nonumber\\
    I_- & = \left| \int_{-1}^1 dv_z G(v_z) \Theta[-G(v_z)] \right|,
\end{align}
respectively, where $\Theta$ is the Heaviside theta function.
For the sake of convenience, in the rest of the paper we call the $v_z$ range over which the above integral is smaller (larger) as the ``small'' (``large'') side, and use $v_z^<$ ($v_z^>$) to denote that range.

Based on the observation from the numerical simulations, a complete flavor equilibration is approximately achieved on the small side on a coarse-grained sense.
A general description on the asymptotic survival probability within two-flavor oscillations\footnote{In the three-flavor case where $\nu_\mu$ and $\nu_\tau$ are indistinguishable, the expression is $P_{ee}^{\rm 3f}(v_z) = 1-4[1-P_{ee}^{\rm 2f}(v_z)]/3$.} can be given as
\begin{equation}
    P_{ee}^{\rm 2f}(v_z) =
    \begin{cases}
        \frac{1}{2} & {\rm for~}v_z^<, \\
        P_{ee}(v_z) & {\rm for~}v_z^>,
    \end{cases}
\end{equation}
where the distributions on the large side $P_{ee}(v_z)$ can be formulated by a chosen analytical prescription. 
Below, based on the same assumption on the small side, we will describe both previously formulated prescriptions and the improved ones proposed in this paper.

\subsection{Prescriptions with abrupt transition}
\label{sec:abrupt}
It was suggested in Refs.~\cite{zaizen2023simple,bhattacharyya2021fast} to use a boxlike expression, whose spatially averaged survival probabilities $P_{ee}(v_z)$ are constant in $v_z^>$, to describe the asymptotic distribution for neutrino survival probabilities. 
Assuming that the small side undergoes a complete flavor equilibration, the conservation of total $\nu$ELN requires on the large side that
\begin{equation}
    P_{ee}(v_z) = 1-\frac{I_<}{2 I_>},
    \label{eq:express_box}
\end{equation}
where $I_<=\min(I_-,I_+)$ and $I_>=\max(I_-,I_+)$.

Another prescription assumes that $P_{ee}(v_z)$ is linear in $v_z$ on the large side \cite{bhattacharyya2021fast,bhattacharyya2022elaborating}:
\begin{equation}
    P_{ee}(v_z) =
    \frac{1}{2}+\frac{I_>-I_<}{4I_>}[ 1\mp v_c\mp \frac{3}{2}(v_c^2-1)v_z ],
    \label{eq:express_linear}
\end{equation}
where the sign $-$ ($+$) denotes that $v_z=1$ is on the large (small) side.
Furthermore, one can follow the same method of Refs.~\cite{bhattacharyya2021fast,bhattacharyya2022elaborating} to  include the second-order quadratic Legendre polynomial so that $P_{ee}(v_z)$ is quadratic in $v_z$. 
This adds an additional term to Eq.~\eqref{eq:express_linear} and results in
\begin{align}
    P_{ee}(v_z) = 
    \frac{1}{2}+\frac{I_>-I_<}{4I_>} [ & 1\mp v_c \mp \frac{3}{2}(v_c^2-1)v_z \nonumber\\
    &  \mp \frac{5}{4}(v_c^3-v_c)(3v_z^2-1) ].
    \label{eq:express_quadratic}
\end{align}
Both linear and quadratic prescriptions ensure $P_{ee}(v_z)\geq 1/2$ on the large side to avoid introducing an additional zero crossing in the asymptotic state.
However, it is important to note that Eqs.~\eqref{eq:express_linear} and \eqref{eq:express_quadratic} both do not guarantee the conservation of total $\nu$ELN or the constraint that $P_{ee}(v_z)\leq 1$ on the large side.

\subsection{Prescriptions with continuous transition}
None of the above prescriptions ensure a continuous transition at the zero crossing $v_c$, which can lead to an artificial discontinuity in the final asymptotic angular distributions $\tilde g_{\nu_e}$ and $\tilde g_{\bar\nu_e}$.
To avoid this, we propose a new prescription for the large side as
\begin{equation}\label{eq:express_continuous}
    P_{ee}(v_z) = 1-\frac{1}{2}h(|v_z-v_c|/a),
\end{equation}
where $h(x)$ is a $v_z$-dependent function that monotonically decreases from 1 to 0 when $x$ increases from 0 to infinity.
We try three different double-power laws for $h(x)$ as $(x^2+1)^{-1/2}$, $(x^2+1)^{-1}$, and $(x+1)^{-2}$, denoted as power-1/2, -1, and -2, respectively.
In addition, we take one more exponential function $h(x)=\exp(-x)$.
For any choice of $h(x)$, the coefficient $a$ can be numerically solved using the Newton-Raphson method for the following equation
\begin{equation}
    I_< = \int_{v_z^>} dv_z G(v_z) h(|v_z-v_c|/a),
    \label{eq:constraint_a}
\end{equation}
which can be derived based on the $\nu$ELN conservation.

Because the right-hand side of Eq.~\eqref{eq:constraint_a} is monotonic in $a$, an interpolation method can also be used to effectively solve the coefficient $a$ in practice.
For a set of $a$, $\{a_i\}$, including several finite positive $a_i$ values as well as $a_i=0^+$ and $a_i=\infty$, the  corresponding values for $\Gamma(a_i) = (I_>)^{-1}\int_{v_z^>} dv_z G(v_z) h(|v_z-v_c|/a_i)$ can be calculated covering a range from 0 to 1.
For a given $I_</I_>$, one can find the interval defined by a pair of adjacent values $a_i$ and $a_{i+1}$ where $I_</I_>$ is sandwiched by $\Gamma(a_i)$ and $\Gamma(a_{i+1})$. 
Then, the asymptotic distribution can be interpolated as
\begin{align}
  P_{ee}(v_z) = 1-\frac{1}{2} \left[\right. & \gamma h(|v_z-v_c|/a_i) \nonumber\\
  & \left. + (1-\gamma) h(|v_z-v_c|/a_{i+1}) \right]
\end{align}
with $\gamma = |\Gamma(a_{i+1})-I_</I_>|/|\Gamma(a_{i+1})-\Gamma(a_{i})|$.
We note that the total $\nu$ELN is also conserved in this practical scheme.
Although this procedure can be applied to any of the four above schemes with continuous transitions, we 
demonstrate its practicability in this work by taking $a_i=\{0^+, 0.04, 0.2, 1, \infty \}$ based on the power-1/2 prescription and denote this practical scheme as power-1/2-i. 

We elaborate further here on the above choice of the monotonic function $h(x)$ on the large side. 
Based on the observation from simulation results, the unstable eigenmode that grows the fastest in the linear regime usually has larger amplitude at the small side in $v_z$. 
When evolving into the nonlinear regime, it often results in more flavor conversion closer to the small side.
Since the unstable eigenmode has a continuous distribution in $v_z$, this implies that for a $v_z$ that is farther away from the small side, it generally experiences less flavor conversion.
As a result, when the system relaxes to the quasistationary state through kinematic decoherence, the spatially averaged survival probability keeps the memory of the continuous transition, which leads to a typically larger value of $P_{ee}(v_z)$ closer to the small side.

\section{Results}
\label{sec:results}
In this section, we assess the performance of various analytical prescriptions in predicting the asymptotic distributions of survival probabilities and the relevant angular moments.
We use superscripts ``sim'' and ``pre'' to distinguish those quantities from the simulations and prescriptions, respectively.
We will compare eight prescriptions including boxlike, linear, quadratic, power-1/2, power-1/2-i, power-1, power-2, and exponential ones described in Sec.~\ref{sec:prescriptions} in the following analysis.

\subsection{Two representative conditions}
\label{sec:two_representatives}
\begin{figure*}[hbt!]
	\centering
		\includegraphics[width=0.66\columnwidth]{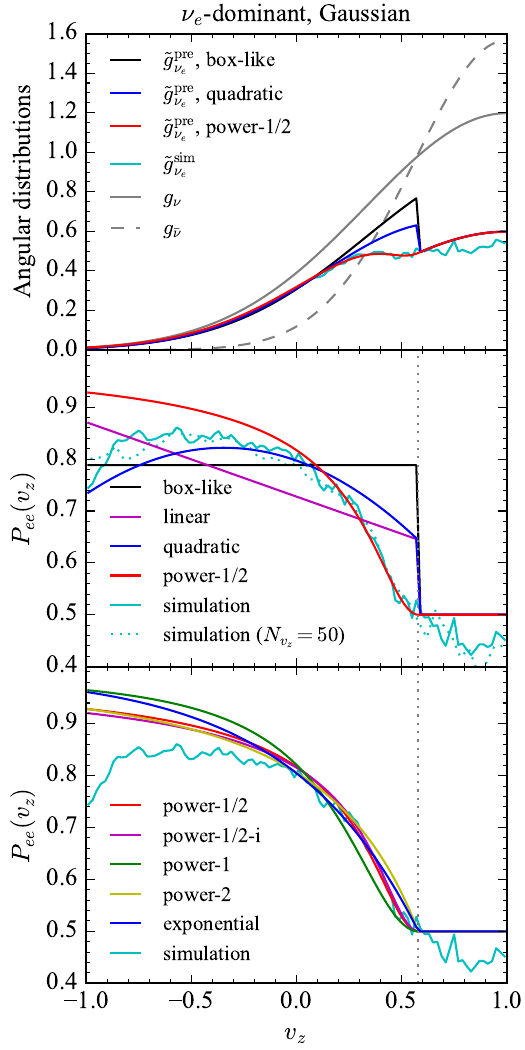}
        \llap{\parbox[b]{1.6in}{\small (a)\\\rule{0ex}{4.1in}}}\hspace{-0.04in}
        \llap{\parbox[b]{1.6in}{\small (b)\\\rule{0ex}{2.75in}}}\hspace{-0.04in}
        \llap{\parbox[b]{1.6in}{\small (c)\\\rule{0ex}{1.4in}}}\hspace{-0.04in}
		\includegraphics[width=0.66\columnwidth]{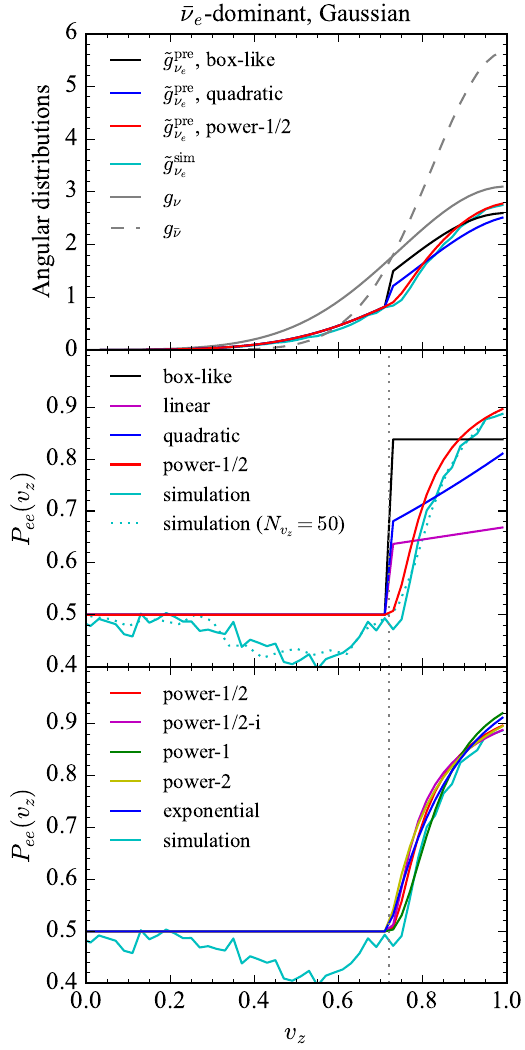}
        \llap{\parbox[b]{1.6in}{\small (d)\\\rule{0ex}{4.1in}}}\hspace{-0.04in}
        \llap{\parbox[b]{1.6in}{\small (e)\\\rule{0ex}{2.75in}}}\hspace{-0.04in}
        \llap{\parbox[b]{1.6in}{\small (f)\\\rule{0ex}{1.4in}}}\hspace{-0.04in}
		\includegraphics[width=0.66\columnwidth]{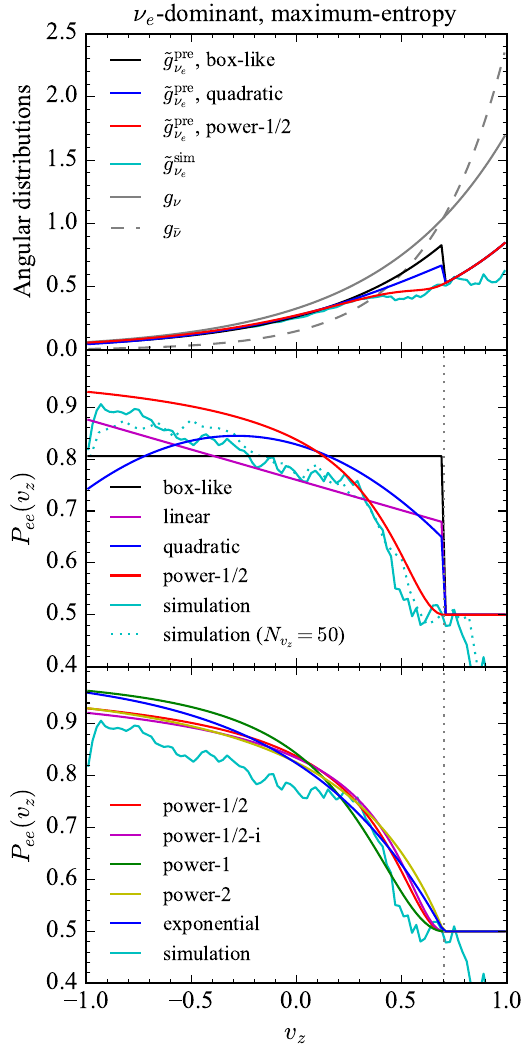}
        \llap{\parbox[b]{1.6in}{\small (g)\\\rule{0ex}{4.1in}}}\hspace{-0.04in}
        \llap{\parbox[b]{1.6in}{\small (h)\\\rule{0ex}{2.75in}}}\hspace{-0.04in}
        \llap{\parbox[b]{1.6in}{\small (i)\\\rule{0ex}{1.4in}}}\hspace{-0.04in}
	\caption{\label{fig:comparison}Angular distributions (a),(d),(g) and survival probabilities (b)-(c),(e)-(f),(h)-(i) with the initial distributions parametrized by the Gaussian type.
    (a)--(c) [(d)--(f)] Smaller (larger) initial $\bar\nu_e$ zeroth moment than $\nu_e$ as described in Sec.~\ref{sec:two_representatives}. 
    (g)--(i) Same $\nu_e$-dominant moments as in (a)--(c) but with distributions characterized by the maximum-entropy type. 
    The gray vertical line marks the zero crossing at $v_c$.
    Note that (d)--(f) only show the $v_z$ range from 0 to 1 for clarity.
 } 
\end{figure*}

To illustrate some general behaviors of those prescriptions, we show in Fig.~\ref{fig:comparison} the angular distributions and asymptotic survival probabilities $P_{ee}(v_z)$ for two representative conditions obtained with simulations as well as those from the analytical prescriptions.
For $P_{ee}(v_z)$, all eight different prescriptions are shown in the plots while for the angular distribution, we only show analytical results derived using the boxlike [Eq.~\eqref{eq:express_box}], quadratic, [Eq.~\eqref{eq:express_quadratic}], and the power-1/2 of Eq.~\eqref{eq:express_continuous}.
The first typical condition represents a system dominated by electron neutrinos with $(I_{\bar\nu},F_\nu,F_{\bar\nu})\approx(0.87,0.47,0.65)$ while the second one is dominated by antineutrinos with $(I_{\bar\nu},F_\nu,F_{\bar\nu})\approx(1.27,0.79,0.86)$.\footnote{Those two conditions are provided as ID 58 and 32, respectively, in the dataset \cite{dataset_fast_asymptotic}.}
Figures~\ref{fig:comparison}(a)--\ref{fig:comparison}(c) [\ref{fig:comparison}(d)--\ref{fig:comparison}(f)] show results obtained in the $\nu_e$ ($\bar\nu_e$)-dominant condition with the initial angular distributions $g_\nu$ and $g_{\bar\nu}$ parametrized by the Gaussian-type distribution.  
Because the chosen parameter in the $\bar\nu_e$-dominant condition indicates more forward-peaked distributions than that in the $\nu_e$-dominant condition, we only show the $v_z$ range from 0 to 1 in Figs.~\ref{fig:comparison}(d)--\ref{fig:comparison}(f).
For the $\nu_e$-dominant case, we also show in Figs.~\ref{fig:comparison}(g)--\ref{fig:comparison}(i) results obtained with $g_\nu$ and $g_{\bar\nu}$ parametrized by the maximum-entropy distribution.

For both conditions, $G(v_z)>0$ for $v<v_c$ because $F_\nu<F_{\bar\nu}$, i.e., $g_{\nu}$ is less forward peaked than $g_{\bar\nu}$.
This range is the large side when $I_{\bar\nu}<1$ ($\nu_e$-dominated case) and only undergoes incomplete conversions toward flavor equilibration as shown in Figs.~\ref{fig:comparison}(a)--\ref{fig:comparison}(c).
When $I_{\bar\nu}>1$, this range becomes the small side and reaches approximate flavor equilibration shown in Figs.~\ref{fig:comparison}(d)--\ref{fig:comparison}(f). 
Obviously, approximate flavor equilibration and incomplete flavor conversions are  obtained in the range $v_z<v_c$ for cases with $I_{\bar\nu}<1$ and $I_{\bar\nu}>1$, correspondingly.
In addition, we also show that very similar results are obtained with $N_{v_z}=50$ when compared to those obtained with $N_z=100$ in Figs.~\ref{fig:comparison}(b) and \ref{fig:comparison}(e).

For the analytical prescriptions, Fig.~\ref{fig:comparison} shows that large differences on the large side exist between results obtained using Eqs.~\eqref{eq:express_box}--\eqref{eq:express_continuous}.
However, using Eq.~\eqref{eq:express_continuous} with different double-power or exponential functions $h(x)$ generally gives rise to similar outcomes.
For $P_{ee}(v_z)$ with the $\nu_e$-dominant condition ($I_{\bar\nu_e}<1$), both the boxlike and linear prescriptions contain relatively large deviations $\sim 0.1$--$0.3$ from simulation outcome near the zero crossing $v_c$.
The quadratic prescription matches better with the simulation result in the range of  $v_z\lesssim 0.1$. 
However, the deviation near $v_c$ is similarly large as with the linear prescription because the second-order Legendre polynomial has a small contribution near $v_z=\sqrt{3}/3\approx 0.57$.
As mentioned already in Sec.~\ref{sec:prescriptions}, the large deviations around $v_c$ are related to the inherit discontinuities in these prescriptions.

Significant improvements are obtained near $v_c$ when using prescriptions with continuous transitions.
For all continuously transitioning cases with either the double-power law or the exponential function, the survival probabilities increase from 0.5 at the small side to larger values as $v_z$ decreases without any discontinuity.
More specifically, the power-2 and exponential prescriptions contain first-order discontinuity at $v_c$, while the other two prescriptions are first-order continuous at $v_c$, which leads to slightly more flat $P_{ee}$ at $v_z\lesssim v_c$ and therefore more flavor conversions.
All these prescriptions have small deviations of $\lesssim 0.04$ from the simulation result in the range of $0<v_z<v_c$, with both power-1/2 and power-1/2-i schemes showing the best agreement.
For $v_z<0$, larger deviations up to $\sim 0.15$ appear for all of them, particularly for the power-1 and exponential prescriptions.
However, in terms of the angular distributions, because the initial $g_{\nu(\bar\nu)}$ are forward peaked, the deviations at negative $v_z$ only result in negligible differences in the asymptotic angular distribution $\tilde g_{\nu_e}(v_z)$, as shown in Fig.~\ref{fig:comparison}(a).
On the other hand, the abrupt transition of $P_{ee}$ at $v_c$ for all discontinuous prescriptions lead to artificial peaks in $\tilde g_{\nu_e}(v_z)$ with an obvious discontinuity at $v_c$.
We note here that if one plots the oscillated $\nu$ELN distributions as in Ref.~\cite{zaizen2023simple}, these discontinuities at $v_c$ will disappear because $G(v_c)=0$ and the deviations around $v_c$ will appear small. 
Nevertheless, the physical angular distributions $g_{\nu}$ and $g_{\bar\nu}$ are generally nonzero there so that this feature of discontinuity can hardly be avoided.

With the $\bar\nu_e$-dominant condition ($I_{\bar\nu}>1$), the large side ranges from $v_z=v_c$ to $v_z=1$ with the zero crossing $v_c\approx 0.72$ as shown in Fig.~\ref{fig:comparison}(d).
Similar to the $\nu_e$-dominant case, taking the boxlike, linear, or quadratic prescription also results in larger deviations from simulation outcome than taking the continuous prescriptions.
Interestingly, $P_{ee}(v_z)$ with the quadratic prescription appears nearly linear in this case. 
This is because the additional quadratic Legendre contribution is derived based on the whole $v_z$ range, which results in a large linear term compared to the quadratic term for the narrower $v_z>v_c$ range where we apply the prescription.
For all the continuous prescriptions, the agreements in $P_{ee}$ with the simulation results appear to be even better than the $\nu_e$-dominant case.

For the $\nu_e$-dominant case with the same parameter set $(I_{\bar\nu},F_\nu,F_{\bar\nu})\approx(0.87,0.47,0.65)$ but taking the initial $g_{\nu(\bar\nu)}$ given by the maximum-entropy distributions, Figs.~\ref{fig:comparison}(g)--\ref{fig:comparison}(i) show that the resulting asymptotic distributions are qualitatively similar to those obtained with Gaussian function discussed above. 
Compared to the Gaussian case, the zero crossing $v_c$ is shifted from $\approx 0.6$ to 0.7, but the small side remains at $v_z>v_c$.
Here, different analytical prescriptions for the large side with abrupt transitions at $v_c$ also show similarly large differences in $P_{ee}(v_z)$, while those formulated with continuous transitions result in similar $P_{ee}(v_z)$ that match better with the numerical result. 
Note that here the asymptotic $P_{ee}(v_z)$ obtained with simulations do contain some noticeable differences from the Gaussian case shown in Fig.~\ref{fig:comparison}(b).
This can affect the comparison of different analytical prescriptions to simulation outcome. 
For instance, the linear prescription now performs better than the quadratic one in $v_z<0$ as shown in Fig.~\ref{fig:comparison}(h). 
Also, the region where approximate flavor equilibration is achieved is extended to $v_z\approx 0.5$ below $v_c$, around which the power-1 prescription fits the simulation result better.

Looking at the small side, although all schemes assume the same flavor equilibration with $P_{ee}=0.5$ as an approximation, we observe an interesting phenomenon that slight overconversions are possible for some parameter sets with either types of initial distributions.
For example, $P_{ee}$ can be $\sim 0.4$ for $v_z\gtrsim 0.8$ for the $\nu_e$-dominant case with the initial maximum-entropy distribution, and $0.5\lesssim v_z\lesssim 0.6$ for the $\bar\nu_e$-dominant case with the initial Gaussian distribution, independent of the choice of angular resolutions.
The overconversions may result from some specific unstable eigenmodes when fast instability develops from the linear to nonlinear regime (see, e.g., \cite{wu2021collective,bhattacharyya2022elaborating}).
Although it leads to nonzero asymptotic $\nu$ELN on the small side, it does not lead to an additional spectrum crossing and fast flavor instability in the asymptotic state, because the nonzero $\nu$ELN there has the same sign as in the large side.
However, the presence of the overconversions can result in systematic biases in the predictions of the boxlike prescription as well as those continuous ones due to the imposed constraint from the conservation of $\nu$ELN. 
Since there are more flavor conversions than the equilibration on the small side due to the overconversions, the $\nu$ELN conservation then implies that there will also be more flavor conversions on the large side obtained by simulations than results derived with those analytical prescriptions, as shown in Fig.~\ref{fig:comparison}. 

\subsection{Overall performance}
Going beyond the explicit comparisons based on only a few examples, we further evaluate the overall performance of each analytical prescription for all parameter sets by calculating several useful error quantities, including the root mean square errors for $\tilde g_{\nu_e}(v_z)$ over the entire $v_z$ range as well as over the large side only, and the differences for the first two angular moments. 
We write those error quantities in terms of $\nu_e$ explicitly as
\begin{align}
    E(\tilde g_{\nu_e}) & = \frac{1}{2}\left[ \int_{-1}^1 dv_z \left|\tilde g^{\rm pre}_{\nu_e}(v_z)-\tilde g^{\rm sim}_{\nu_e}(v_z) \right|^2 \right]^{1/2}, \nonumber\\
    E(\tilde g_{\nu_e}^>) & = \frac{\left[ \int_{v_z^>} dv_z \left|\tilde g^{\rm pre}_{\nu_e}(v_z)-\tilde g^{\rm sim}_{\nu_e}(v_z) \right|^2 \right]^{1/2}}{\int_{v_z^>} dv_z}, \nonumber\\
    E(\tilde I_{\nu_e}) & = | \tilde I^{\rm pre}_{\nu_e}-\tilde I^{\rm sim}_{\nu_e} |, \nonumber\\
    E(\tilde J_{\nu_e}) & = | \tilde J^{\rm pre}_{\nu_e}-\tilde J^{\rm sim}_{\nu_e} |.
    \label{eq:error}
\end{align}
The error $E(\tilde g_{\nu_e}^>)$ is evaluated excluding the contribution from the small side because the same flavor equilibration is assumed in all prescriptions.
One can replace all subscripts of $\nu_e$ by $\bar\nu_e$ for the corresponding errors in the antineutrino sector.

We notice that some $\nu$ELN distributions in our sample have very ``shallow'' zero crossings, i.e., small ratios of $(I_</I_>)\ll 1$.
For these cases, their zero crossings are close to $v_z=-1$ or 1. 
As a result, nearly no flavor conversion occurs on the large side due to the $\nu$ELN conservation, similar to the conditions found in large radii of a CCSN \cite{abbar2022suppression}.
These cases can be empirically classified as with no flavor conversion and hence are not included in the performance comparison.
After excluding these shallow distributions with the ratio $(I_</I_>)<10^{-2}$, the numbers of parameter sets are reduced to $N^{\rm set}=7479$ and 7162 for the Gaussian and maximum-entropy types, respectively.

\begin{figure*}[hbt!]
	\centering
		\includegraphics[width=.32\textwidth]{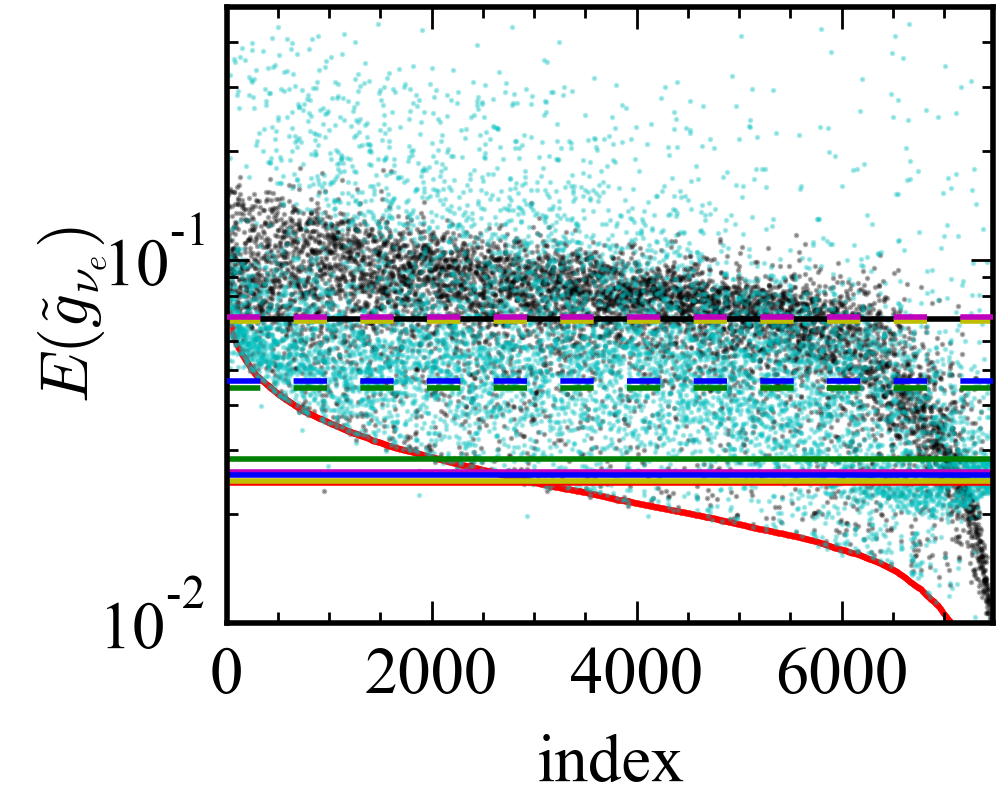}
        \llap{\parbox[b]{0.5in}{\small (a)\\\rule{0ex}{1.6in}}}\hspace{-0.04in}
		\includegraphics[width=.32\textwidth]{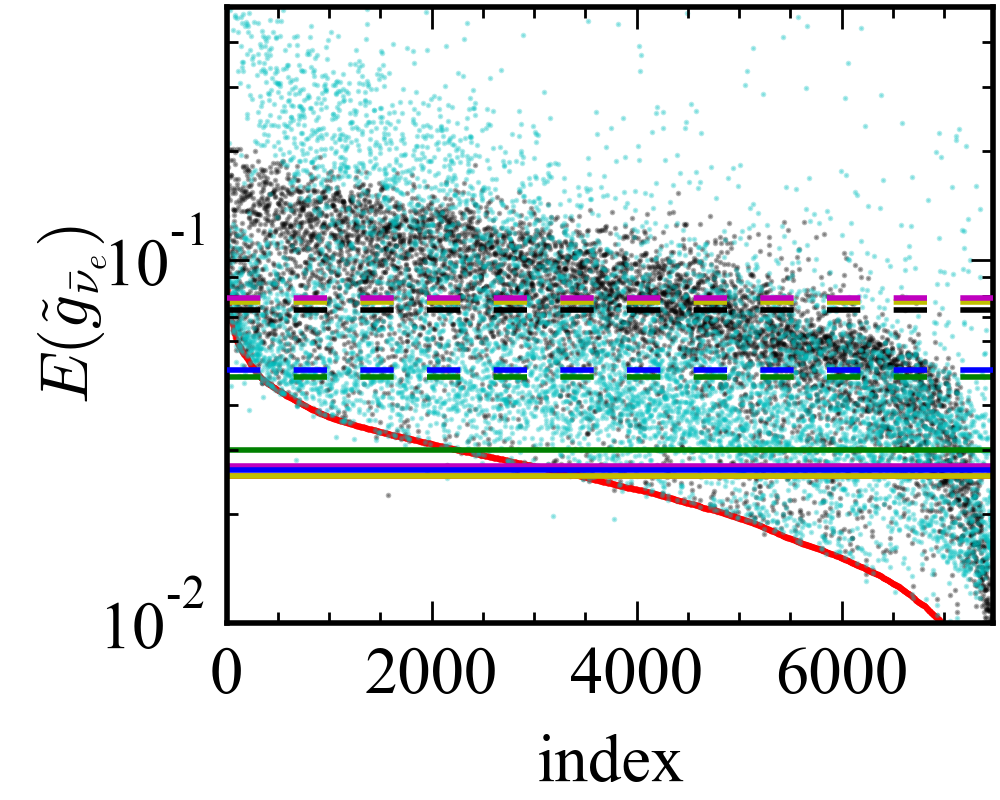}
        \llap{\parbox[b]{0.5in}{\small (b)\\\rule{0ex}{1.6in}}}\hspace{-0.04in}
		\includegraphics[width=.32\textwidth]{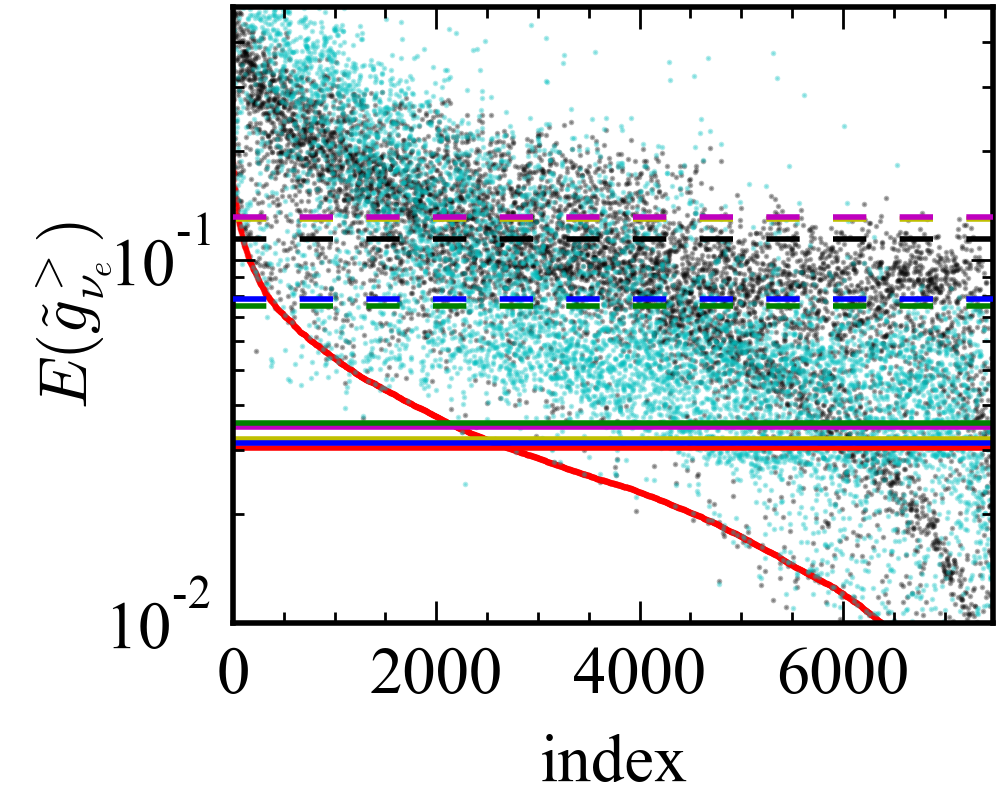}
        \llap{\parbox[b]{0.5in}{\small (c)\\\rule{0ex}{1.6in}}}\hspace{-0.04in}
		\includegraphics[width=.32\textwidth]{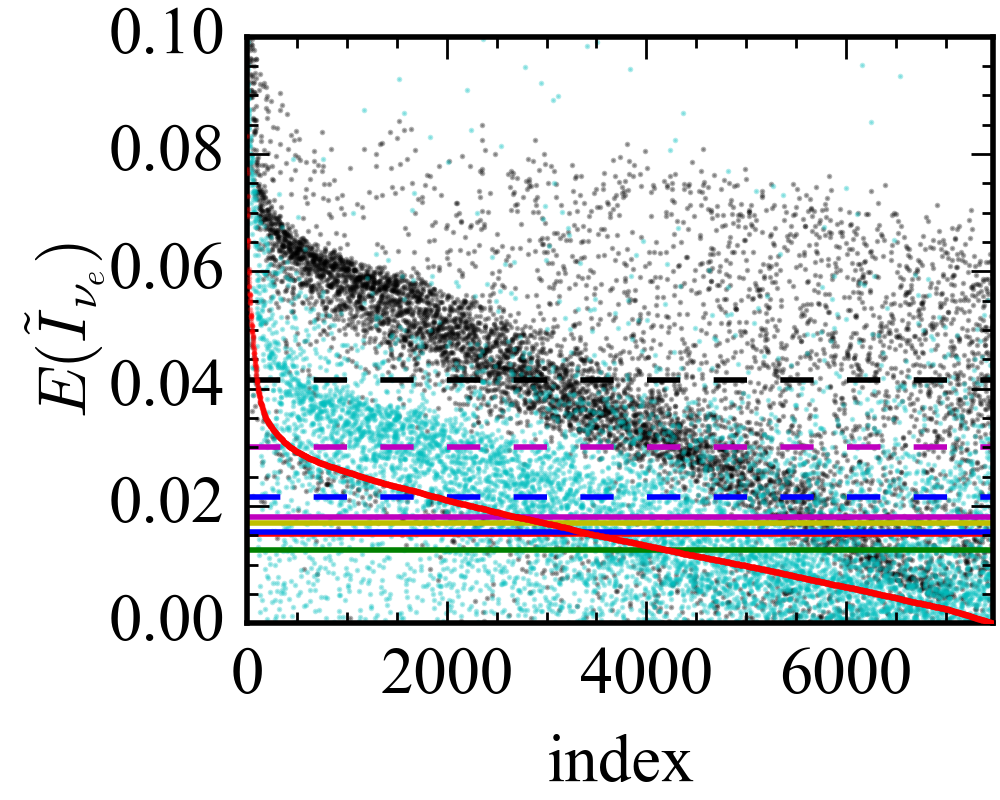}
        \llap{\parbox[b]{0.5in}{\small (d)\\\rule{0ex}{1.5in}}}\hspace{-0.04in}
		\includegraphics[width=.32\textwidth]{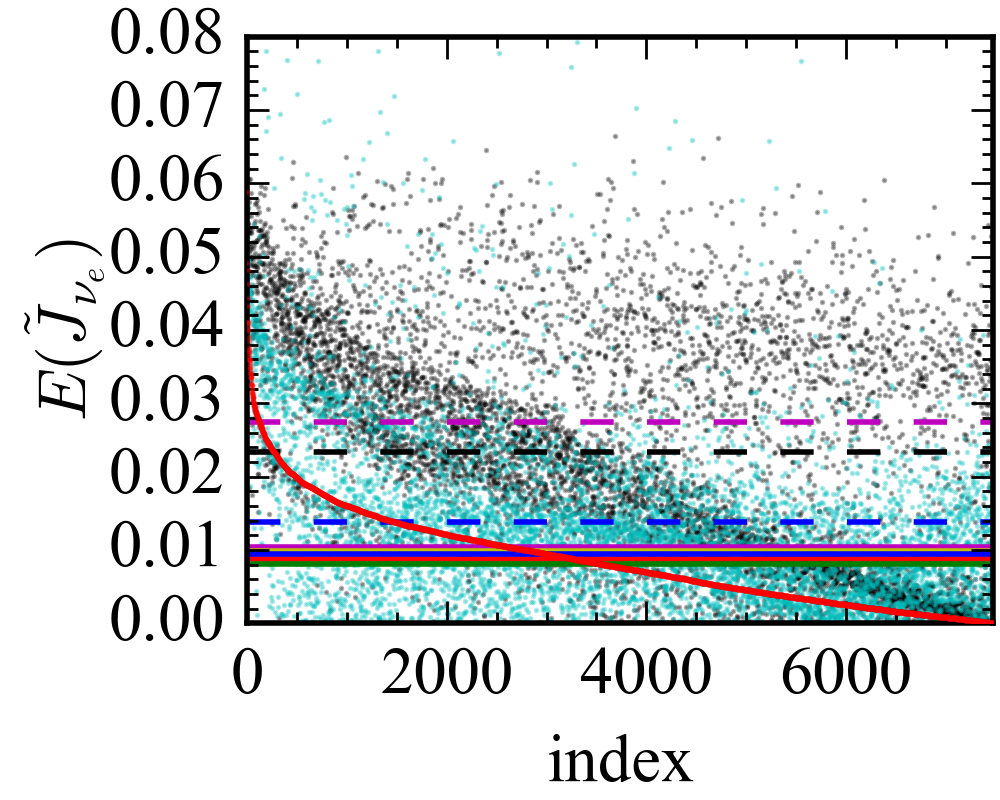}
        \llap{\parbox[b]{0.5in}{\small (e)\\\rule{0ex}{1.5in}}}\hspace{-0.04in}
		\includegraphics[width=.32\textwidth]{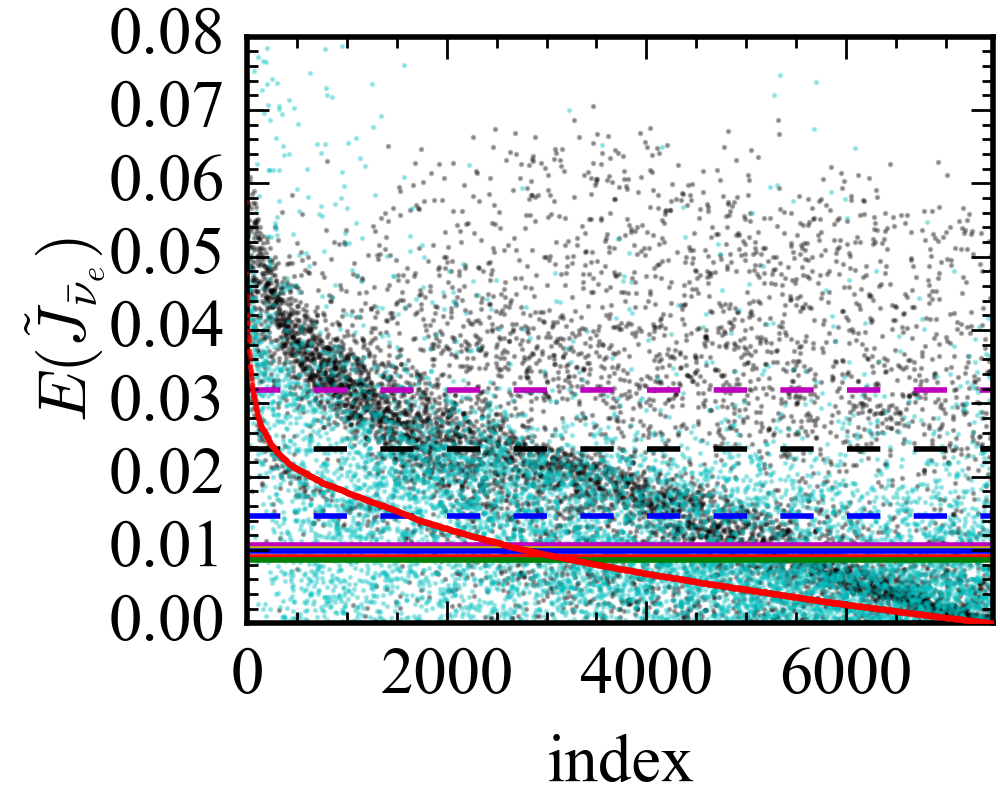}
        \llap{\parbox[b]{0.5in}{\small (f)\\\rule{0ex}{1.5in}}}\hspace{-0.04in}
	\caption{\label{fig:error}Error quantities of distributions (a)--(c) and angular moments (d)--(f) defined in Eqs.~\eqref{eq:error} for the boxlike (black dots), quadratic (cyan dots), and power-1/2 (red dots) prescriptions. 
    For each panel, the indices of the parameter sets are sorted so that the errors obtained with the power-1/2 scheme follow a descending order, so that the red dots form a red curve. 
    The Gaussian type is adopted for the initial angular distributions. The horizontal lines show the arithmetic mean errors $\mathcal E$ over the whole sets as defined in Eq.~\eqref{eq:error_arithmetic} for the boxlike (black dashed), linear (magenta dashed), quadratic (blue dashed), power-1/2 (red solid), power-1/2-i (magenta solid), power-1 (green solid), power-2 (yellow solid), and exponential (blue solid) prescriptions, respectively. 
    Note that additional cases where the truncation of $P_{ee}=1$ is implemented for the linear and quadratic prescriptions are shown in (a)--(c) by the yellow and green dashed lines, respectively.}
\end{figure*}

Figure~\ref{fig:error} shows the error quantities for all nonshallow distributions of the Gaussian type with the boxlike, quadratic, and power-1/2 prescriptions for all parameter sets.  
The indices for each panel are numbered such that the corresponding error quantities obtained with the power-1/2 prescription decrease with increasing index numbers for the sake of clearer presentation.
The top panels show that the power-1/2 prescriptions clearly outperform the boxlike and quadratic predictions for the distributional errors $E(\tilde g_{\nu_e})$, $E(\tilde g_{\bar\nu_e})$ and $E(\tilde g_{\nu_e}^>)$.  
Most of the $E(\tilde g_{\nu_e})$ and $E(\tilde g_{\bar\nu_e})$ with the boxlike prescription sit around $\sim 0.08$. 
Comparatively, the power-1/2 scheme provide improvement for these two errors by up to $\sim 75\%$ for parameter sets with indices $\sim 5000$.
When taking the quadratic prescription, although it generally gives rise to smaller errors compared to the boxlike scheme, these two errors have larger variations with errors as large as 0.5 for some parameter sets.
For $E(\tilde g_{\nu_e}^>)$, similar features hold except that now there exist large variations in errors for all three prescriptions. 
The underlying reason is that the common contribution from the small side is not included when computing $E(\tilde g_{\nu_e}^>)$. 

The distributions of errors in moments $E(\tilde I_{\nu_e})$, $E(\tilde J_{\nu_e})$, and $E(\tilde J_{\bar\nu_e})$ show somewhat different and interesting behaviors.
Both the boxlike and quadratic prescriptions show large variations.
For some cases, they can perform better than the power-1/2 one in moments despite their worse performance in distributional errors discussed above. 
The seemly contradictory result is related to the cancellation of positive and negative contributions of errors in Eq.~\eqref{eq:error} as well as the overconversions and systematic biases discussed in Sec.~\ref{sec:two_representatives}.
For example, the asymptotic zeroth moments $\tilde I_{\nu_e}$ in the antineutrino-dominant case presented in Figs.~\ref{fig:comparison}(d)--\ref{fig:comparison}(f) are $\approx 0.673$, 0.681, and 0.707 for the simulation, quadratic, and power-1/2 prescriptions, respectively. 
Although clearly the $\tilde g(\nu_e)$ obtained with the power-1/2 prescription resembles better the $g(\nu_e)$ from the simulation than with the quadratic scheme, the latter results in smaller $E(\tilde I_{\nu_e})$ due to the cancellation of contributions from the integration range of $0.7<v_z<0.85$ and from $v_z>0.85$. 
Such a cancellation does not happen with the power-1/2 prescription where the $\nu$ELN conservation is imposed, because it predicts larger values of $P_{ee}(v_z)$ and $\tilde g_{\nu_e}$ than simulation values on both the small side and the large side due to the overconversions on the small side (see discussions in Sec.~\ref{sec:two_representatives}).
If we have taken the amount of overconversion contribution $\int_{v_z^<} dv_z g_\nu (1/2-P_{ee}^{\rm sim})\approx 0.014$ into account for the power-1/2 scheme, it will lead to a reduced $\tilde I_{\nu_e}=0.679$, which will be better than 0.681 obtained with the quadratic scheme.

We further calculate the arithmetic mean error for each prescription
\begin{equation}
    \mathcal E = \frac{1}{N^{\rm set}}\sum_i^{N^{\rm set}} E_i,
    \label{eq:error_arithmetic}
\end{equation}
where $i$ sums over the whole ensemble of parameter sets.
Those mean errors are shown by the horizontal lines in Fig.~\ref{fig:error} with values listed in Table~\ref{tab:error}.
The prescriptions with abrupt and continuous transitions at $v_c$ are represented by the dashed and solid lines, respectively.
In addition, to prevent $P_{ee}(v_z)$ from exceeding the unity in the linear and quadratic prescriptions as discussed in Sec.~\ref{sec:abrupt}, both Eqs.~\eqref{eq:express_linear} and \eqref{eq:express_quadratic} are replaced by $\min[1, P_{ee}(v_z)]$, and the corresponding arithmetic errors are shown in Figs.~\ref{fig:error}(a)--\ref{fig:error}(c) as the yellow and green dashed lines.
Statistically, the truncation at $P_{ee}=1$ presents negligible improvement on the mean error.
Both the boxlike and linear prescriptions yield the largest errors for all measures.
Comparatively, the quadratic prescription provides visible improvements while all other prescriptions with continuous transition at $v_c$ further reduce the mean errors and show similar performances in general.
It is noteworthy that the power-1/2-i scheme interpolating with five points for the coefficient $a$ in Eq.~\eqref{eq:constraint_a} has errors merely $\sim 5\%$--20\% greater than those with the power-1/2 prescription.

\begingroup
\begin{table}[t]
\begin{ruledtabular}
    \caption{\label{tab:error}Arithmetic mean errors $\mathcal E/10^{-2}$ for all eight prescriptions. The Gaussian type is adopted for the initial angular distributions.}
    \centering
    \begin{tabular}{c cccc cccc}
       Prescriptions & $\tilde g_{\nu_e}$ & $\tilde g_{\bar\nu_e}$ & $\tilde g_{\nu_e}^>$ & $\tilde g_{\bar\nu_e}^>$ & $\tilde I_{\nu_e}$ & $\tilde I_{\bar\nu_e}$ & $\tilde J_{\nu_e}$ & $\tilde J_{\bar\nu_e}$  \\\hline
        Boxlike    & 6.89 & 7.30 & 11.4 & 12.4 & 4.14 & 4.14 & 2.34 & 2.38 \\
        Linear      & 6.97 & 7.86 & 13.2 & 15.3 & 3.01 & 3.46 & 2.74 & 3.18 \\
        Quadratic   & 4.65 & 4.99 & 7.81 & 8.70 & 2.15 & 2.12 & 1.39 & 1.46 \\
        Power-1/2   & 2.44 & 2.55 & 3.04 & 3.46 & 1.52 & 1.52 & 0.87 & 0.91 \\
        Power-1/2-i & 2.62 & 2.72 & 3.47 & 3.87 & 1.82 & 1.82 & 1.04 & 1.07 \\
        Power-1     & 2.83 & 3.01 & 3.56 & 4.11 & 1.25 & 1.25 & 0.81 & 0.86 \\
        Power-2     & 2.46 & 2.55 & 3.21 & 3.57 & 1.71 & 1.71 & 0.98 & 1.01 \\
        Exponential & 2.57 & 2.65 & 3.15 & 3.50 & 1.55 & 1.55 & 0.95 & 0.99 
    \end{tabular}
\end{ruledtabular}
\end{table}
\endgroup

To further check whether the ranking of the arithmetic averages may be affected by specific outlier parameter sets with large error, we calculate one more metric $\mathcal R$, which is the fraction of the best performance for each type of errors $E$ in a subset of prescriptions.
As the average errors from prescriptions with abrupt and continuous transitions are clearly separated into two groups, we use two subsets: subset A includes boxlike, linear, quadratic, and power-1/2 prescriptions, and subset B includes four prescriptions with continuous transition at $v_c$.
They are compared in Table~\ref{tab:error_rank} for cases with initial Gaussian angular distributions.
Consistent with the previous analysis, the power-1/2 prescription has the best performance predominantly in $\approx 98\%$ of all samples among the prescription subset A.
In terms of the predictions for first two moments, the linear and quadratic schemes can perform better for $\approx 10\%$--30\% of the parameter sets, while the power-1/2 prescription performs better for $\approx 45\%$--60\% of all samples.
In subset B, the power-1/2 and power-2 prescriptions share similar best performance percentages $\approx 30\%$ for $\tilde g_{\nu_e}$ and $\tilde g_{\bar\nu_e}$, while the power-1/2 and exponential ones are slightly worse.
With regards to the first two moments, the power-1 prescription dominates and give rises to best values of $\mathcal R\approx 63\%$--73\%.

\begingroup
\begin{table}[t]
\begin{ruledtabular}
    \caption{\label{tab:error_rank}
    Best performance fraction $\mathcal R$ in percentage in two subsets of prescriptions. The Gaussian type is adopted for the initial angular distributions. The total number of parameter sets is $N^{\rm set}=7479$.}
    \centering
    \begin{tabular}{c cc cccc}
        Subset A & $\tilde g_{\nu_e}$ & $\tilde g_{\bar\nu_e}$ & $\tilde I_{\nu_e}$ & $\tilde I_{\bar\nu_e}$ & $\tilde J_{\nu_e}$ & $\tilde J_{\bar\nu_e}$ \\\hline
        Boxlike    & 0.5  & 0.5  & 6.3  & 6.4  & 7.4  & 7.8 \\
        Linear      & 0.5  & 0.4  & 28.0 & 25.0 & 10.3 & 9.7 \\
        Quadratic   & 1.4  & 1.6  & 22.2 & 24.3 & 22.2 & 22.9 \\
        Power-1/2   & 97.7 & 97.5 & 43.5 & 44.3 & 60.1 & 59.6 \\\hline\hline
        Subset B & $\tilde g_{\nu_e}$ & $\tilde g_{\bar\nu_e}$ & $\tilde I_{\nu_e}$ & $\tilde I_{\bar\nu_e}$ & $\tilde J_{\nu_e}$ & $\tilde J_{\bar\nu_e}$ \\\hline
        Power-1/2   & 30.5 & 29.2 & 5.5  & 5.5  & 10.3 & 12.1 \\
        Power-1     & 15.8 & 14.2 & 72.9 & 72.9 & 66.2 & 63.3 \\
        Power-2     & 30.6 & 31.9 & 15.2 & 15.2 & 17.6 & 17.9 \\
        Exponential & 23.1 & 24.8 & 6.4  & 6.4  & 6.0  & 6.7  
    \end{tabular}
\end{ruledtabular}
\end{table}
\endgroup

Most of the results discussed above do not change when taking the maximum-entropy type of presumed angular distributions. 
To avoid repetition, we only show in Table~\ref{tab:error_rank_maxent} the corresponding best performance fraction $\mathcal R$ obtained here for the maximum-entropy type. 
There, the power-1/2 prescription still performs best in subset A, and the power-1 prescription has the largest $\mathcal R\sim 80\%$ for the angular moments in subset B.
The minor difference is that the power-1 prescription also outperforms the power-1/2 and power-2 schemes in $\tilde g_{\nu_e}$ and $\tilde g_{\bar\nu_e}$. 

\begingroup
\begin{table}[t]
\begin{ruledtabular}
    \caption{\label{tab:error_rank_maxent}Same as in Table~\ref{tab:error_rank} except that the maximum-entropy type is adopted for the initial angular distributions. The total number of parameter sets is $N^{\rm set}=7162$.}
    \centering
    \begin{tabular}{c cc cccc}
        Subset A & $\tilde g_{\nu_e}$ & $\tilde g_{\bar\nu_e}$ & $\tilde I_{\nu_e}$ & $\tilde I_{\bar\nu_e}$ & $\tilde J_{\nu_e}$ & $\tilde J_{\bar\nu_e}$ \\\hline
        Boxlike    & 0.3  & 0.3  & 3.5  & 3.4  & 3.9  & 3.8  \\
        Linear      & 0.2  & 0.3  & 19.0 & 16.4 & 7.1  & 6.5  \\
        Quadratic   & 1.0  & 0.8  & 16.8 & 17.0 & 18.1 & 17.2 \\
        Power-1/2   & 98.5 & 98.7 & 60.6 & 63.1 & 70.8 & 72.5 \\\hline\hline
        Subset B & $\tilde g_{\nu_e}$ & $\tilde g_{\bar\nu_e}$ & $\tilde I_{\nu_e}$ & $\tilde I_{\bar\nu_e}$ & $\tilde J_{\nu_e}$ & $\tilde J_{\bar\nu_e}$ \\\hline
        Power-1/2   & 25.2 & 25.1 & 1.1  & 1.1  & 2.2  & 2.3  \\
        Power-1     & 55.0 & 54.6 & 80.0 & 80.0 & 78.9 & 78.4 \\
        Power-2     & 9.8  & 10.4 & 7.3  & 7.3  & 7.6  & 7.7  \\
        Exponential & 10.0 & 9.9  & 11.6 & 11.6 & 11.3 & 11.6 
    \end{tabular}
\end{ruledtabular}
\end{table}
\endgroup

\subsection{Dependence in parameter space}
\begin{figure}[hbt!]
	\centering
		\includegraphics[width=\columnwidth]{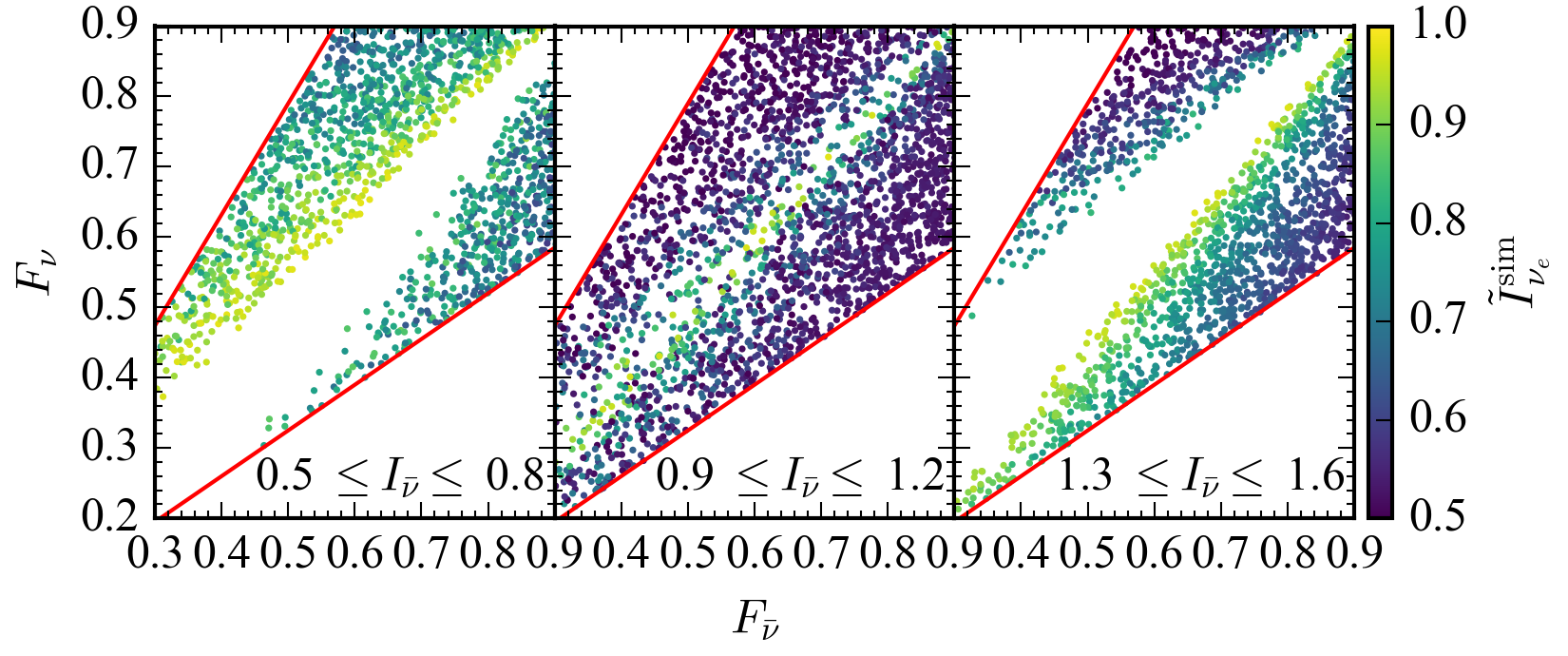}
    \hspace{-0.04in}\llap{\parbox[b]{5.85in}{\small (a)\\\rule{0ex}{1.15in}}}
    \hspace{-0.04in}\llap{\parbox[b]{4.1 in}{\small (b)\\\rule{0ex}{1.15in}}}
    \hspace{-0.04in}\llap{\parbox[b]{2.35in}{\small (c)\\\rule{0ex}{1.15in}}}
    \includegraphics[width=\columnwidth]{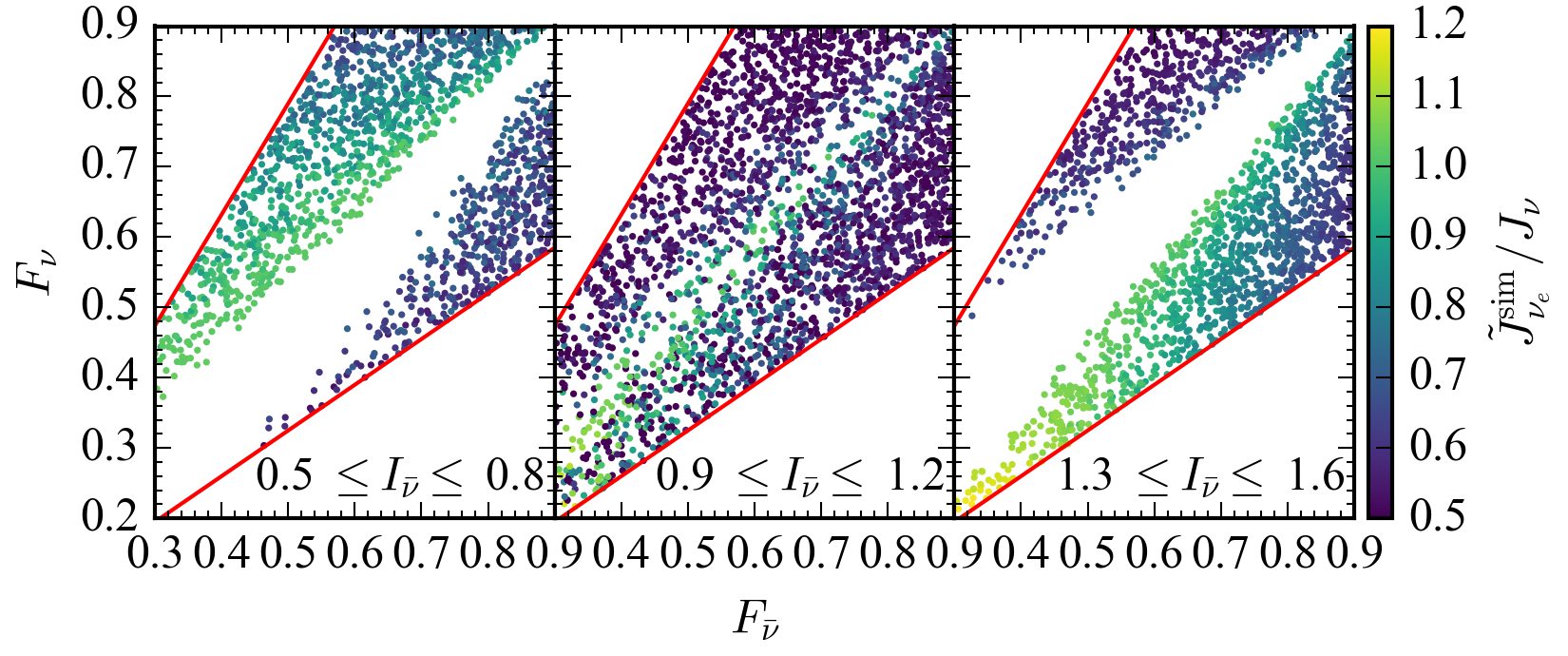}
    \hspace{-0.04in}\llap{\parbox[b]{5.85in}{\small (d)\\\rule{0ex}{1.15in}}}
    \hspace{-0.04in}\llap{\parbox[b]{4.1 in}{\small (e)\\\rule{0ex}{1.15in}}}
    \hspace{-0.04in}\llap{\parbox[b]{2.35in}{\small (f)\\\rule{0ex}{1.15in}}}
	\caption{\label{fig:momentdomain}Parameter dependence of the asymptotic values of the zeroth (a)--(c) and first (d)--(f) moments obtained by simulations. The regions between the two red lines are where we generate our parameter sets. The Gaussian-type initial angular distributions are adopted for this figure.}
\end{figure}

\begin{figure}[hbt!]
	\centering
		\includegraphics[width=\columnwidth]{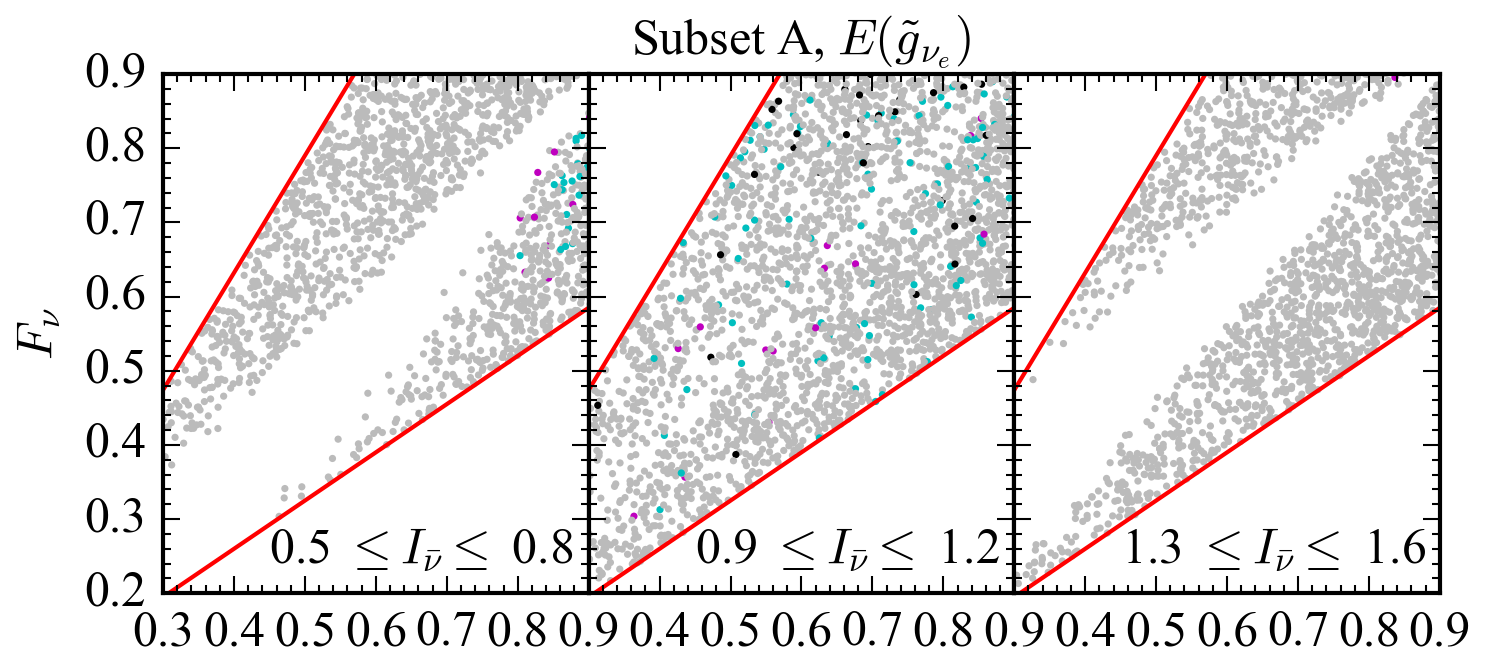}
    \hspace{-0.04in}\llap{\parbox[b]{5.75in}{\small (a)\\\rule{0ex}{1.1in}}}
    \hspace{-0.04in}\llap{\parbox[b]{3.8 in}{\small (b)\\\rule{0ex}{1.1in}}}
    \hspace{-0.04in}\llap{\parbox[b]{1.85in}{\small (c)\\\rule{0ex}{1.1in}}}
    \includegraphics[width=\columnwidth]{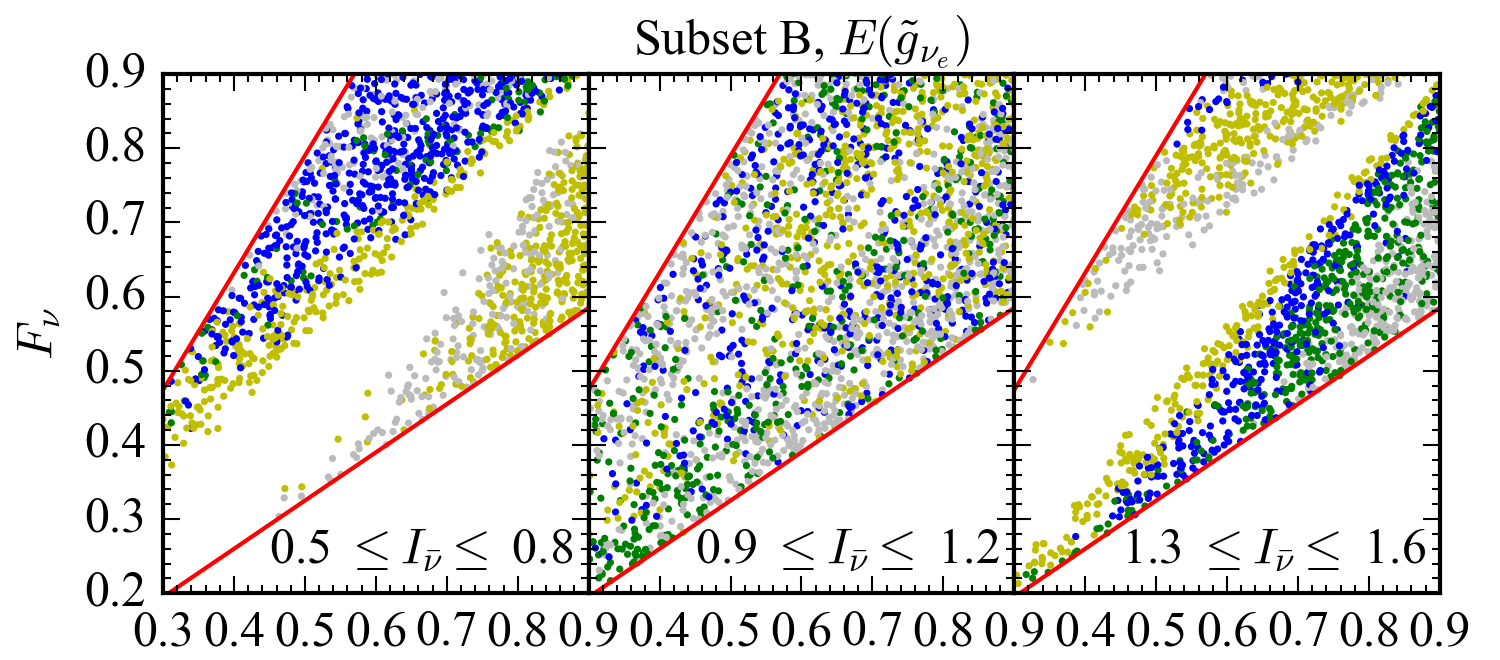}
    \hspace{-0.04in}\llap{\parbox[b]{5.75in}{\small (d)\\\rule{0ex}{1.1in}}}
    \hspace{-0.04in}\llap{\parbox[b]{3.8 in}{\small (e)\\\rule{0ex}{1.1in}}}
    \hspace{-0.04in}\llap{\parbox[b]{1.85in}{\small (f)\\\rule{0ex}{1.1in}}}
    \includegraphics[width=\columnwidth]{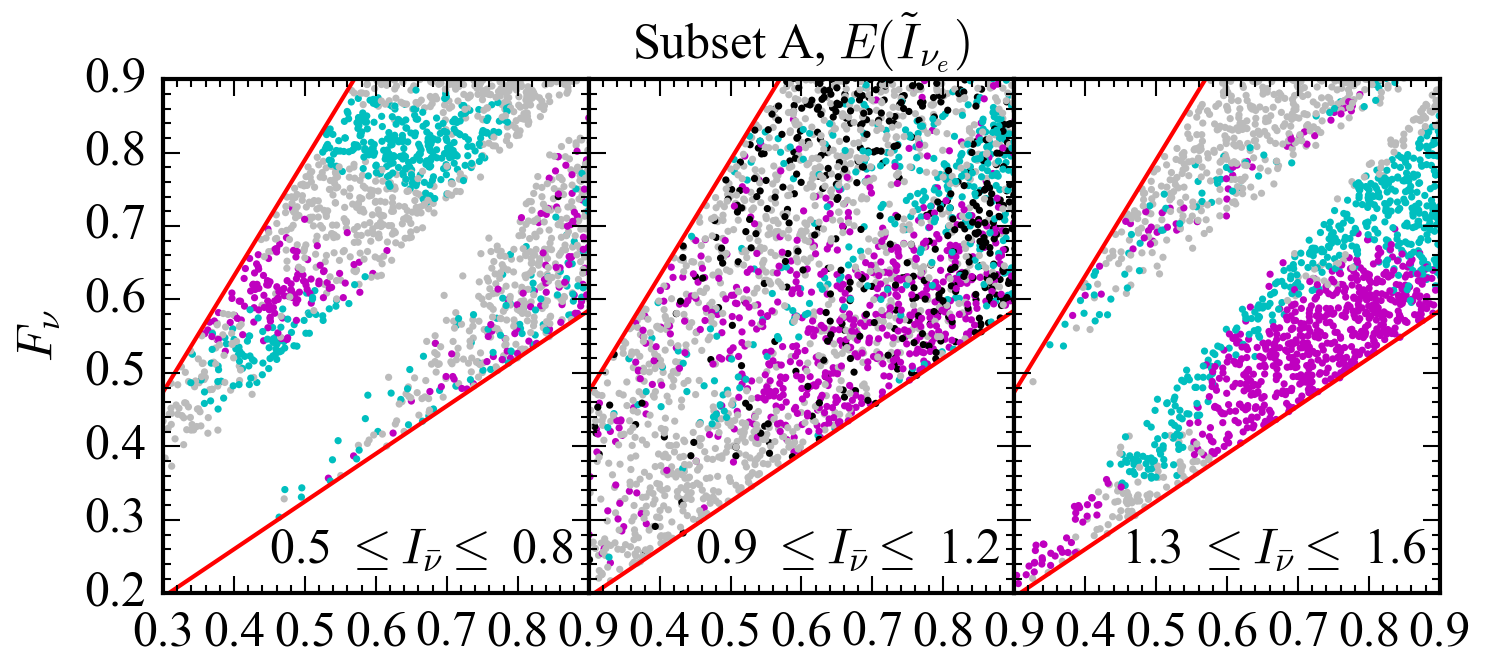}
    \hspace{-0.04in}\llap{\parbox[b]{5.75in}{\small (g)\\\rule{0ex}{1.1in}}}
    \hspace{-0.04in}\llap{\parbox[b]{3.8 in}{\small (h)\\\rule{0ex}{1.1in}}}
    \hspace{-0.04in}\llap{\parbox[b]{1.85in}{\small (i)\\\rule{0ex}{1.1in}}}
    \includegraphics[width=\columnwidth]{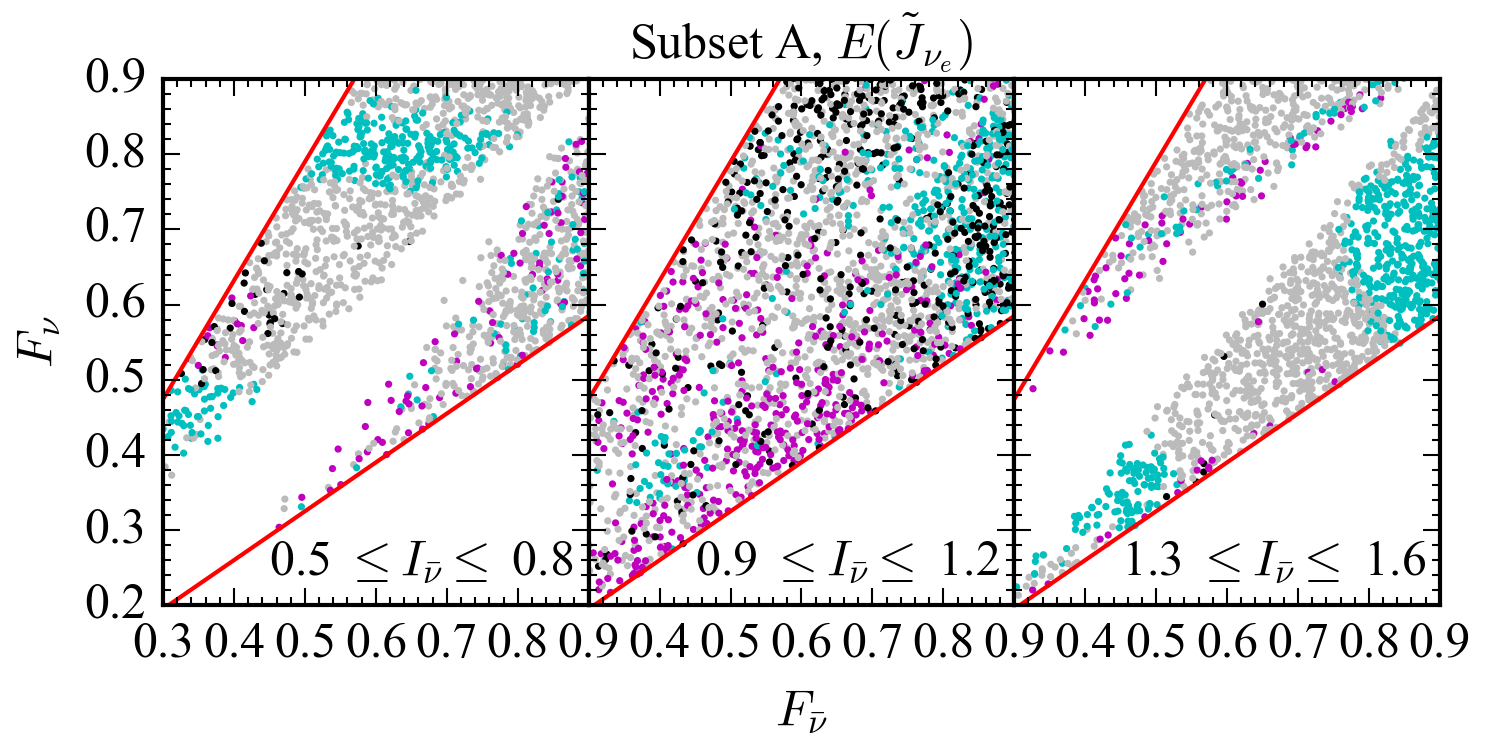}
    \hspace{-0.04in}\llap{\parbox[b]{5.75in}{\small (j)\\\rule{0ex}{1.3in}}}
    \hspace{-0.04in}\llap{\parbox[b]{3.8 in}{\small (k)\\\rule{0ex}{1.3in}}}
    \hspace{-0.04in}\llap{\parbox[b]{1.85in}{\small (l)\\\rule{0ex}{1.3in}}}
	\caption{\label{fig:domain}Parameter dependence of the best performed prescriptions for $E(\tilde g_{\nu_e})$ with the prescription subset A in (a)--(c), $E(\tilde g_{\nu_e})$ with the prescription subset B in (d)--(f), $E(\tilde I_{\nu_e})$ with subset A in (g)--(i), and $E(\tilde J_{\nu_e})$ with subset A in (j)--(l). Colors of the dots represent different types of prescriptions including the boxlike (black), linear (magenta), quadratic (cyan), power-1/2 (gray), power-1 (green), power-2 (yellow), and exponential (blue) ones. 
    The two red lines delineate the region where we generate our parameter sets as in Fig.~\ref{fig:momentdomain}. The Gaussian type is adopted for the initial angular distributions.}
\end{figure}

The errors in Fig.~\ref{fig:error} are ranked regardless of the shape of the distributions or the moments. 
To gain a better understanding on how a specific prescription works better in certain range of the explored parameter space, we examine the dependence of the simulation outcome and the errors associated with each analytical prescriptions in this section. 
For this purpose, we show the qualitative features of the asymptotic values of moments from the simulations, and the evaluated errors with different prescription in the parameter space of $I_{\bar\nu}$, $F_\nu$, and $F_{\bar\nu}$ in Figs.~\ref{fig:momentdomain} and \ref{fig:domain}, respectively.
In both figures, the empty diagonal region in each panel indicates where the fast flavor instability does not exist.
It is more likely to have fast instabilities when the flux factors are significantly different from each other or when $I_{\bar\nu}$ is closer to $I_{\nu}=1$.

When $I_{\bar\nu}\approx 1$, and $F_{\nu}$ greatly deviates from $F_{\bar\nu}$, the zero crossing typically appears in rather central part of the $v_z$ range. 
As a result, near flavor equilibration ($\tilde I_{\nu_e}^{\rm sim}\simeq \tilde J_{\nu_e}^{\rm sim}/J_\nu\simeq 0.5$) for both zeroth and first moments is achieved in most parameter sets as shown by the darker regions in Figs.~\ref{fig:momentdomain}(b)~and~\ref{fig:momentdomain}(e). 
When $F_{\nu}\approx F_{\bar\nu}$, the initial shape of distributions for $\nu_e$ and $\bar\nu_e$ are so similar that the zero crossing of the $\nu$ELN is close to either $v_z=-1$ or $v_z=1$, which leads to incomplete flavor conversion on the large side even with $I_{\bar\nu}\approx 1$.
Such a similar trend also applies to the cases with a large asymmetry of zeroth moments between $\nu_e$ and $\bar\nu_e$ for larger or smaller $I_{\bar\nu_e}$ shown in Figs.~\ref{fig:momentdomain}(a), \ref{fig:momentdomain}(c), \ref{fig:momentdomain}(d), and \ref{fig:momentdomain}(f). 
Closer to the central blank regions of these panels, both the $\tilde I_{\nu_e}^{\rm sim}$ and $\tilde J_{\nu_e}^{\rm sim}/J_\nu$ deviate from 0.5 systematically.
Nearly complete flavor conversion can only happen when $F_{\nu}$ and $F_{\bar\nu}$ differ significantly. 

For most of the parameter sets, the changes of the zeroth and first moments are correlated.
However, unlike the zeroth moment of electron flavor neutrinos that can only decrease assuming no $\nu_x$ in the initial state, the first moment after the FFC can be larger than the initial value. 
For example, at the bottom-left corner of Fig.~\ref{fig:momentdomain}(f) [$1.3\leq I_{\bar\nu}\leq 1.6$] where $F_\nu\sim 0.25$ and $F_{\bar\nu}\sim 0.35$, the ratio $\tilde J_{\nu_e}^{\rm sim}/J_{\nu}$ can be $\approx 1.2$.
This is because the flavor conversion occurs mostly in the range of the backpropagating neutrinos with $v_z<0$, which contribute a non-negligible amount to the moments.

Let us now look at how different analytical prescriptions work in different regions of the moment space. 
Figures~\ref{fig:domain}(a)--\ref{fig:domain}(c) show that, independent of $I_{\bar\nu}$, $F_\nu$, and $F_{\bar\nu}$, the power-1/2 prescription has universally the best performance in the subset A in predicting the asymptotic distribution $\tilde g_{\nu_e}$; see also Table~\ref{tab:error_rank}.  
When considering the subset B shown in Figs.~\ref{fig:domain}(d)--\ref{fig:domain}(f), different prescriptions occupy visibly different parameter space for $0.5\leq I_{\bar\nu}\leq 0.8$ and $1.3\leq I_{\bar\nu}\leq 1.6$ for providing least errors in $E(\tilde g_{\nu_e})$. 
For instance, Fig.~\ref{fig:domain}(f) shows that with the antineutrino-dominant condition $1.3\leq I_{\bar\nu}\leq 1.6$ and $F_\nu<F_{\bar\nu}$, the best prescription gradually transitions from the power-2 to the exponential followed by the power-1 and then power-1/2 types, as the flux factors increase. 
In the same plot but at the corner with $F_\nu>F_{\bar\nu}$, the best prescription transitions from the power-1/2 type to the power-2 type followed by the exponential one.
Interestingly, there does not appear to be any specific prescription that predominately provides the least distributional error $E(\tilde g_{\nu_e})$ in any part of the parameter space with $0.9\leq I_{\bar\nu}\leq 1.2$, indicated by the mixed colors in Fig.~\ref{fig:domain}(e).

Figures~\ref{fig:domain}(g)--\ref{fig:domain}(l) display the best performance prescriptions for $E(\tilde I_{\nu_e})$ and $E(\tilde J_{\nu_e})$ within the subset~A. 
Figures~\ref{fig:domain}(g)--\ref{fig:domain}(i) show somewhat similar domainlike patterns regarding the best performing prescription for $E(\tilde I_{\nu_e})$.  
Although the power-1/2 scheme still outperforms other abrupt prescriptions in a large fraction of the parameter space as indicated by Table~\ref{tab:error_rank}, 
the linear and quadratic prescriptions can outperform the power-1/2 in some particular parameter regions.
For example, the linear prescription has the best performance when $0.5\leq I_{\bar\nu}\leq 0.8$, $F_\nu\sim 0.6$, and $F_{\bar\nu}\sim 0.5$, or when $1.3\leq I_{\bar\nu}\leq 1.6$, $F_\nu\sim 0.4$--0.6, and $F_{\bar\nu}\sim 0.7$--0.8.
The quadratic prescription has the best performance when $0.5\leq I_{\bar\nu}\leq 0.8$, $F_\nu\sim 0.8$, and $F_{\bar\nu}\sim 0.6$, or when $1.3\leq I_{\bar\nu}\leq 1.6$, $F_\nu\sim 0.7$, and $F_{\bar\nu}\sim 0.8$.  
As for the boxlike prescription, it only provides better performance for certain parameter sets that are sparsely distributed in Fig.~\ref{fig:domain}(h) with $0.9\leq I_{\bar\nu}\leq 1.2$.

Comparing Figs.~\ref{fig:domain}(j)--\ref{fig:domain}(l) showing the best prescription distribution for $E(\tilde J_{\nu_e})$ to Figs.~\ref{fig:domain}(g)--\ref{fig:domain}(i), it shows that the distribution of the best-performance domains can vary when considering different moments. 
For instance, although for the quadratic prescription it performs the best in similar parameter space for both $E(\tilde I_{\nu_e})$ and $E(\tilde J_{\nu_e})$, in the region with $1.3\leq I_{\bar\nu}\leq 1.6$, $F_\nu\sim 0.4$--0.6, and $F_{\bar\nu}\sim 0.7$--0.8, the power-1/2 replaces the linear prescription as the best prescription in subset A for $E(\tilde J_{\nu_e})$. 

\subsection{Effects of heavy-lepton flavor neutrinos}
\label{sec:nux}
In the previous discussions the heavy-lepton flavor neutrinos are not taken into consideration in the initial angular distributions because the $\nu$ELN distribution is unchanged if the same amount of $\nu_x$ and $\bar\nu_x$ distributions are assumed.
As a result, the evolution and asymptotic distribution for survival probabilities are expected to be the same, although the asymptotic angular distributions $\tilde g_{\nu_e}$ and $\tilde g_{\bar\nu_e}$ can be affected.

\begin{figure}[hbt!]
  \centering
    \includegraphics[width=\columnwidth]{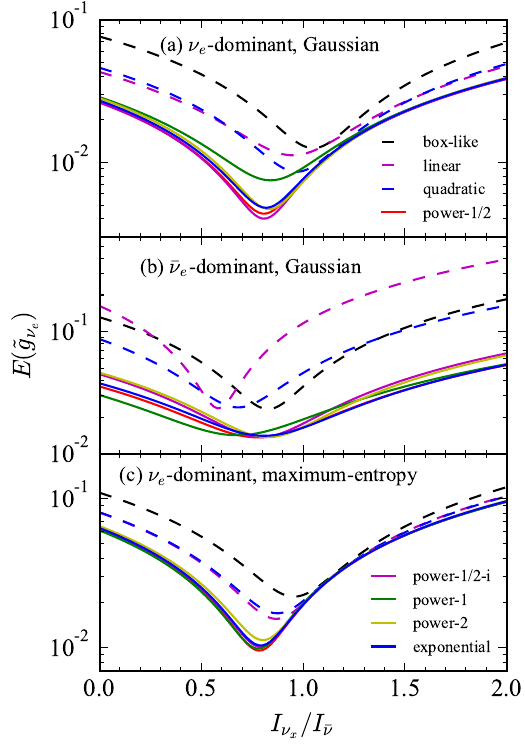}
  \caption{\label{fig:nux}Distributional errors $E(\tilde g_{\nu_e})$ as functions of $I_{\nu_x}/ I_{\bar\nu}$ for eight prescriptions in the initial conditions of Gaussian (a),(c) and maximum-entropy (b) distributions.}
\end{figure}

Because the inclusion of heavy-lepton flavor neutrinos introduces more dimensions in the parameter space, we do not perform detailed analysis here for the performance evaluation.
Instead, we provide a specific example below to illustrate how to evaluate its impact on the errors obtained in earlier sections, which can be generally applied by postprocessing the dataset that we released.
Assuming both $\nu_x$ and $\bar\nu_x$ have the same flux factor as $\bar\nu_e$ initially for simplicity, their angular distributions can be characterized by the zeroth moment $I_{\nu_x}$ as
\begin{equation}
    g_{\nu_x}(v_z) = g_{\bar\nu_x}(v_z) = \frac{I_{\nu_x}}{I_{\bar\nu}} g_{\bar\nu}(v_z).
\end{equation}
With the asymptotic survival probability unaffected, the final distribution for $\nu_e$ now becomes
\begin{equation}\label{eq:nux}
    \tilde g_{\nu_e}(v_z) = g_{\nu}(v_z) P_{ee}(v_z) + \frac{I_{\nu_x}}{I_{\bar\nu}} g_{\bar\nu}(v_z) [1-P_{ee}(v_z)],
\end{equation}
for both the simulated $\tilde g_{\nu_e}^{\rm sim}$ and predicted $\tilde g_{\nu_e}^{\rm pre}$.
In Fig.~\ref{fig:nux}, we show the distributional errors $E(\tilde g_{\nu_e})$ as a function of $I_{\nu_x}/I_{\bar\nu}$ computed based on Eq.~\eqref{eq:nux} 
for the two conditions considered in Sec.~\ref{sec:two_representatives} with the initial distributions parametrized by the Gaussian and maximum-entropy functions.

The errors $E(\tilde g_{\nu_e})$ for all prescriptions start to decrease as $I_{\nu_x}$ increases at the beginning.
They reach the minimum at $I_{\nu_x}/I_{\bar\nu}\sim 0.6$--1 and eventually increase again.
The reduction of errors for $I_{\nu_x}\sim I_{\bar\nu}$ is because for more similar distributions of $\nu_e$ and $\nu_x$, less changes to the $\nu_e$ distribution can occur due to the conversion of $\nu_x$ to $\nu_e$. 
Specifically, the minimum locates at $I_{\nu_x}/I_{\bar\nu}$ slightly less than 1 where the crossing between angular distributions of $g_\nu$ and $g_{\nu_x}$ happens at the small side, as this allows one to minimize the systematic error contribution due to the observed overconversions seen in simulations discussed earlier.

Comparing different analytical prescriptions, Figs.~\ref{fig:nux}(a)--\ref{fig:nux}(c) show that those with the continuous transitions at $v_c$ again provide similar errors with different $I_{\nu_x}/I_{\bar\nu}$ and generally perform better than the abrupt prescriptions.
For the abrupt ones, the quadratic prescription still has better performance than the boxlike and linear ones, but can be slightly outperformed, e.g., by the linear scheme at $I_{\nu_x}/I_{\bar\nu}\sim 0.6$ or the boxlike scheme at $I_{\nu_x}/I_{\bar\nu}\sim 0.9$ in the antineutrino-dominant condition with the initial Gaussian type shown in Fig.~\ref{fig:nux}(b).

\section{Discussion and conclusions}
\label{sec:conclusions}
In this paper, we conducted a comprehensive survey over a large sample of initial neutrino angular distributions to investigate the outcome of the asymptotic state of FFC in the periodic 1D-box setup. 
Several thousands of simulations for initial $\nu_e$ and $\bar\nu_e$ angular distributions parametrized by the Gaussian and the maximum-entropy functions that can also be specified by the initial zeroth moment of $\bar\nu_e$, $I_{\bar\nu}$, and flux factors $F_\nu$ and $F_{\bar\nu}$ were performed to times when the systems reach close to the asymptotic states.
These results provide a database for the design of effective treatments so that FFCs can be approximately incorporated into realistic hydrodynamic simulations that include  classical neutrino transport.

We found that in the asymptotic state, flavor conversions on one side of the $\nu$ELN (defined as the small side in this work)  happen in a way that the system evolves toward flavor equilibration to eliminate the zero crossing when averaging over the entire box, as pointed out in several earlier works \cite{bhattacharyya2021fast,wu2021collective,richer2021neutrino}. 
Interestingly, we also found that slight overconversions on the small side in the asymptotic state can happen as a general final outcome of the system, which however, does not introduce new zero crossings.

Assuming flavor equilibration on the small side, we formulated several new analytical prescriptions that aim to improve the existing formulations including the boxlike and the linear prescriptions proposed in Refs.~\cite{bhattacharyya2021fast,bhattacharyya2022elaborating,zaizen2023simple}, which provided analytical formulas to characterize the asymptotic state on the large side of the $\nu$ELN. 
One of our new proposals extends these existing ones and includes the second-order Legendre polynomial correction, resulting in a quadratic velocity dependence on the large side. 
More importantly, to overcome the artificial discontinuity encountered at the zero crossing when using the boxlike, linear, and quadratic expressions, we provided several new prescriptions that continuously connect the flavor conversion probabilities on the small and the large sides while respecting the $\nu$ELN conservation. 

Based on our simulation data, we first compared in detail the asymptotic states predicted by these analytical prescriptions with those obtained numerically for two representative examples.
We then evaluated the overall performance for the entire datasets using several error measures, including the distributional errors, net differences of the first two angular moments, and the fraction of best performance. 
We found that despite the fast that all prescriptions provide reasonable predictions with small distributional errors $\lesssim 0.15$ and angular moment differences $\lesssim 0.05$, 
the prescriptions with continuous transitions at zero crossings systematically outperform those with abrupt transitions.  
Specifically, the quadratic prescription reduces the average errors by $\sim 30\%$--50\% from the boxlike and linear schemes, while all the continuous prescriptions give rise to another factor of $\sim 30\%$--60\% improvement from the quadratic scheme mainly due to the imposed condition continuity around $v_c$. 

There exist certain advantages and disadvantages associated with these prescriptions. 
The evaluation of the boxlike, linear, and quadratic schemes can be directly done with the explicit formulas given in Sec.~\ref{sec:abrupt}. 
For the linear and the extended quadratic schemes, they were derived by adding corrections upon the boxlike prescription; neither the $\nu$ELN conservation nor non-negative transition probability is ensured unless some additional truncation is imposed.
For the continuous prescriptions, it in principle requires extra computational efforts to solve the width coefficient $a$ iteratively.
However, we also demonstrated that one can use the interpolation method to obtain the asymptotic distributions efficiently without sacrificing much the accuracy of their predictive power.
Moreover, if one wants to directly implement these prescriptions with neutrino transport solvers that adopt the discrete-ordinate schemes, taking the abrupt prescriptions will introduce large errors associated with angular distribution discontinuity when numerically evaluating the angular advection, which can be avoided with continuous prescriptions.

We have also discussed the dependence of the outcome on the parameter space and the impact when including non-negligible heavy-lepton neutrinos in the initial condition.
We found that similar conclusions discussed above generally hold for cases including the heavy-lepton flavors---the continuous prescriptions perform better than those abrupt schemes. 
However, we also noted for some parameter space, the linear and quadratic schemes (without $\nu$ELN conservation constraint) in fact give rise to smaller errors in individual angular moment differences than all the continuous prescriptions, due to the accidental cancellation effect when integrating over the distributions.  
For the continuous prescriptions with $\nu$ELN conservation being imposed, the generally obtained flavor overconversions prevent the accidental cancellation to occur and causes a larger systematic bias. 

Several questions remain to be addressed beyond this work and we list a few below. 
Do the overconversions on the small side depend on the periodic boundary conditions?
Will they be suppressed in the presence of collisions?
If not, how can we improve the formulation of the asymptotic state to account for the overconversions? 
Can these prescriptions be applicable to more general scenarios, e.g., cases where the azimuthal symmetry is broken? 
Answering all those questions certainly requires many follow-up studies and will help achieve the ultimate goal of implementing flavor conversions of neutrinos in hydrodynamical simulations of supernovae and neutron-star mergers.\\

The survey dataset for this paper is publicly available from the Zenodo repository \cite{dataset_fast_asymptotic}.


\begin{acknowledgments}
We thank Oliver Just, Gabriel Mart\'{i}nez-Pinedo, and Yong-Zhong Qian for fruitful discussions.
Z.~X., M.-R.~W., and S.~A. are grateful to the Mainz Institute for Theoretical Physics (MITP) of the Cluster of Excellence PRISMA+ (Project ID 39083149) for its hospitality and its partial support during the completion of this work.
Z.~X. acknowledges support of the European Research Council (ERC) under the European Union’s Horizon 2020 research and innovation program (ERC Advanced Grant KILONOVA No.~885281).
M.-R.~W., S.~B., and M.~G. acknowledge support from the National Science and Technology Council, Taiwan under Grants No.~110-2112-M-001-050 and No.~111-2628-M-001-003-MY4, the Academia Sinica under Project No.~AS-CDA-109-M11, and Physics Division, National Center for Theoretical Sciences, Taiwan.
S.~A. was supported by the German Research Foundation (DFG) through the Collaborative Research Centre ``Neutrinos and Dark Matter in Astro- and Particle Physics (NDM),'' Grant No.~SFB-1258, and under Germany’s Excellence Strategy through the Cluster of Excellence ORIGINS EXC-2094-390783311.
The following software was used in this work: \textsc{numpy} \cite{numpy}, \textsc{matplotlib} \cite{matplotlib}, and \textsc{scipy} \cite{scipy}.
\end{acknowledgments}

\bibliography{main.bbl}

\end{document}